\documentclass[12pt,preprint]{aastex}
\shorttitle{NGC7538}
\shortauthors{Sandell et al.}
\slugcomment{Accepted :  Nov 19,  2004}

\begin{document}

\def\arcmin{{$^{\prime}$}}
\def\arcsec{{$^{\prime\prime}$}}
\def\ptsec{$''\mskip-7.6mu.\,$}
\def\psec{$^s\mskip-7.6mu.\,$}
\def\Msun{\,{\rm M$_{\odot}$}}
\def\Lsun{\,{\rm L$_{\odot}$}}
\def\ltsim{$\stackrel{<}{\sim}$}
\def\gtsim{$\stackrel{>}{\sim}$}
\def\jtra#1#2{${\rm J}\!\!=\!\!{#1}\!\!\to\!\!{#2}$}
\def\degr{$^{\circ}$}

\title{Protostars and Outflows in the NGC\,7538 - IRS\,9 Cloud Core}

\author{G\"oran Sandell\altaffilmark{1}, W. M. Goss\altaffilmark{2}, and Melvyn Wright\altaffilmark{3}}

\altaffiltext{1}{SOFIA-USRA, NASA Ames Research Center, MS N211-3,
Moffett Field, CA 94035, U.S.A.}
\email{gsandell@mail.arc.nasa.gov}
\altaffiltext{2}{National Radio Astronomy Observatory, P.O. Box O, Socorro, NM 87801, U.S.A.}
\altaffiltext{3}{Radio Astronomy Laboratory, University of California, Berkeley
601 Campbell Hall, Berkeley, CA 94720, U.S.A.}

\begin{abstract}

New high resolution observations of HCO$^+$ \jtra10, H$^{13}$CN \jtra10,
SO 2$_2\to1_1$, and continuum with BIMA at 3.4~mm show that  the
NGC\,7538 - IRS\,9 cloud core is a site of active ongoing star
formation. Our observations reveal at least three young bipolar
molecular outflows, all $\sim$ 10,000 -- 20,000 years old.  IRS\,9
drives a  bipolar, extreme high velocity outflow observed nearly pole
on. South of IRS\,9 we find a cold, protostellar condensation with a
size of $\sim$ 14\arcsec\ $\times$ 6\arcsec\ with a mass $>$ 250 \Msun.
This is the center of one of the outflows and shows deep, red-shifted
self absorption in HCO$^+$, suggesting that there is a protostar
embedded in the core, still in a phase of active accretion. This source
is not detected in the far infrared, suggesting that the luminosity $<$
10$^4$\Lsun; yet the mass of the outflow is $\sim$ 60 \Msun. The
red-shifted HCO$^+$ self-absorption profiles observed toward the
southern protostar and IRS\,9 predict accretion rates of a few times
10$^{-4}$ to 10$^{-3}$ M$_{\odot}$~yr$^{-1}$. Deep VLA continuum
observations at 3.6 cm show that IRS\,9 coincides with a faint thermal
VLA source, but no other young star in the IRS\,9 region has any
detectable free-free emission at a level of $\sim$ 60 $\mu$Jy at 3.6 cm.
The HCO$^+$ abundance is significantly enhanced in the hot IRS\,9
outflow. A direct comparison of mass estimates from HCO$^+$ and CO for
the well-characterized red-shifted IRS\,9 outflow predicts an HCO$^+$
enhancement of more than a factor of 30, or [HCO$^+$/H$_2$] $\ge$
6~10$^{-8}$.

\end{abstract}


\section{Introduction}

Massive stars form in associations or clusters together with hundreds of
intermediate and low-mass stars \citep{Hillenbrand77}. Although it is
still not conclusively proven that high-mass stars form the same way as
low-mass stars, i.e. with accretion disks supporting accretion and
driving outflows, recent theoretical models show that high accretion
rates and the pressure from surrounding massive cloud cores are
sufficient to overcome radiation pressure from a newly formed high-mass
star, therefore allowing the stellar mass to increase far above the
theoretical limit of 8 \Msun\ \citep{Maeder02,McKee02}. Recent  surveys
of high-mass star formation regions \citep{Henning00,Zhang01} show that
molecular outflows are common in high-mass young stars. Accretion disks,
however, are far more difficult to detect, because high-mass protostars
are more distant and deeply embedded than low-mass stars, and since they
form in clusters or small groups it is difficult to isolate and image
disks around  high-mass protostars, because of confusion and poor
spatial resolution. Yet, there are several results that show rather
convincingly that high-mass protostars are surrounded by rotating
accretion disks \citep{Cesaroni97,Zhang98,Sandell03,Chini04,Beuther04b}.
The mass distribution of protostellar clumps in high-mass star forming
regions also appears to be consistent with the stellar initial mass
function \citep{Beuther04a}, strongly supporting accretion as the
dominant mechanism for forming high-mass stars.

The molecular cloud bordering the large \ion{H}{2} region NGC\,7538 to
the SE is known to contain several centers of active and on-going
high-mass star formation
\citep{Werner79,Kameya89,Kameya90,Davis98,Sandell04}. In the following
we have adopted  the distance to NGC\,7538 as 2.8 kpc
\citep{Crampton78}, which is the most commonly used distance estimate,
although we note that two photometric studies \citep{Beetz76,Moreno86}
place NGC\,7538 at a distance of 2.2 kpc. NGC\,7538\,IRS\,9, henceforth
called IRS\,9, is a deeply embedded cold IR source with a luminosity, L
$\sim$ 4~ 10$^4$ \Lsun. IRS\,9 is located $\sim$ 2\arcmin\ SE of the
well studied Hyper  Compact \ion{H}{2} region IRS\,1. The source is
associated with a large reflection nebula
\citep{Werner79,Eiroa88,Tamura91}, drives a powerful extreme high
velocity outflow \citep{Kameya89,Mitchell91,Hasegawa95}, and shows only
weak free-free emission \citep{Kameya90}, suggesting that it has not yet
developed an \ion{H}{2} region. There are several indications that
IRS\,9 may not be the only young stellar object in the IRS\,9 cloud
core. Deep CCD imaging \citep{Campbell88} in r, i, and z filters
revealed a diffuse nebula $\sim$ 50\arcsec\ W of IRS\,9, which they
identified as an HH object. \citet{Davis98} found several knots of
shocked H$_2$ emission at 2.12 $\mu$m, including a 40\arcsec\ collimated
jet. Neither the HH object nor the H$_2$ jet appear to point back
towards IRS\,9, suggesting that these  are powered by other active young
stellar objects embedded in the IRS\,9 molecular cloud core.

In this paper we present  high angular resolution spectral line and continuum
imaging of NGC\,7538\,IRS\,9 with BIMA at 3.4 mm. We also present deep
continuum imaging at 3.6 cm and additional data obtained from the
VLA archive.  Images in HCO$^+$ \jtra21 show  that IRS\,9 drives a compact
bipolar high velocity outflow observed almost pole on. By re-analyzing
published CO \jtra21 data together with our HCO$^+$ data, we detect at
least two additional molecular outflows. One coincides with the H$_2$
jet and is  highly collimated (axial ratio $\sim$ 9)  in blue-shifted
HCO$^+$ emission. The second outflow is located in the dense cloud core
south of IRS\,9 and crosses the H$_2$ jet.  We derive properties of the
outflows and discuss what this implies for the star formation activity
in the seemingly isolated IRS\,9 cloud core.

\section{Observations and Data Reduction}
\subsection{BIMA Observations}

The observations of NGC\,7538\,IRS\,9 were made in 2003 April and 2004
January using one frequency setting in the C and B-array configuration
of the BIMA array resulting in a synthesized beam of $\sim$6\arcsec. The
correlator was split into four 25 MHz bands giving a velocity
resolution of  $\sim$0.34 km~s$^{-1}$. We observed H$^{13}$CN \jtra10,
SO  2$_2\to1_1$ and NH$_2$D 1$_{11}\to1_{01}$ in the lower sideband of
the first LO and HCO$^+$ \jtra10 in the upper sideband.

The data were reduced and imaged in a standard way using MIRIAD
software \citep{Sault95}. Phase calibration was applied using
observations of the quasar 0102$+$584 at intervals of 30 minutes. The
phase calibrator was observed using an 800 MHz bandwidth for 3 min.
3C454.3 was observed for 10 minutes as a bandpass calibration.  The
flux density scale was based on observations of Mars. The
uncertainty in the absolute amplitude scale $\sim$15 \%, but the
relative amplitude of spectral lines within the same receiver tuning is $\sim$5 \%.

The data were imaged with weighting inversely proportional to the
variance in order to obtain the best signal to noise ratio.  Averaging 4
spectral channels we obtained an RMS noise level of 40 mJy (0.2 K) with
a synthesized beam Full Width Half Maximum (FWHM) of 6\ptsec4 $\times$
4\ptsec9 at a position angle (p.a.) of $-$67\degr\ with a peak sidelobe level
of -7 \%.  Spectral windows which did not contain any significant
spectral line emission were averaged to provide a continuum image with
an RMS $\sim$3.5 mJy~beam$^{-1}$. The continuum emission was subtracted
from the spectral line channels, and the images deconvolved using the
CLEAN algorithm.

\subsection{VLA Observations and Archive Data}

We carried out 3.6~cm continuum observations of NGC\,7538 with the Very
Large Array (VLA) of the National Radio Astronomy Observatory
(NRAO)\footnote{The NRAO is a facility of the National Science Foundation
operated under cooperative agreement by Associated Universities, Inc.}
in the BnA configuration on 2003 October 14. The effective bandwidth was
100~MHz with two circular polarizations. The pointing center was chosen
half way between NGC\,7538 - IRS\,1 and NGC\,7538\,S, which gave us a
good coverage of the whole NGC\,7538 molecular cloud. The on source
integration time was 1.3 hrs. We used 2229$+$5695 as the phase
calibrator and 3C\,48  for flux density calibration. The data were
edited and calibrated using the AIPS software package. We estimate the
absolute calibration accuracy to be $\sim$ 5\%. The synthesized beam
FWHM was 1\ptsec21 $\times$ 0\ptsec47,  (p.a. =
$+$0\degr{}).  The RMS noise in the vicinity of IRS\,9  was $\sim$ 60
$\mu$Jy~beam$^{-1}$.

We have additionally searched the VLA data archive for any observations
which may provide some further insight on the star formation activity in
the IRS\,9 region. We have re-reduced a 6~cm and 18~cm VLA run of
NGC\,7538 in the B array on 1989 April 28\footnote{This is project AK224
referred to by \citet{Kameya90}}. The 6~cm track was a continuum track
using 50~MHz bandwidth and the same phase calibrator and flux density
calibrator (2229$+$5695 \& 3C\,48) as our 3.6~cm observations. The on
source time was 2.7 hrs and the synthesized FWHM was 1\ptsec35 $\times$
1\ptsec09  (p.a.  $-$24\degr{}) with an RMS noise $\sim$ 100
$\mu$Jy~beam$^{-1}$ in the vicinity of IRS\,9. The 18~cm run was
configured for observations of 1665, 1667 and 1720 MHz OH maser emission
with 127 channels and a bandwidth of 781.25 kHz corresponding to a
velocity resolution of 0.55~km~s$^{-1}$.  We only searched for maser
emission in the central half of the band, i.e. over a velocity range of 
$-$90 to $-$30~km~s$^{-1}$. The synthesized beam for these observations
was 4\ptsec2 $\times$ 3\ptsec2 (p.a. $+$32\degr{}) and the RMS noise was
13 mJy~beam$^{-1}$ in the line data. We also found some short H$_2$O
observations carried out on 29 January 2001 in the D to A array. The
synthesized beam for these observations was 2\ptsec36 $\times$ 0\ptsec38
(p.a. 0\degr{}).  The bandwidth for these observations was 3.125 MHz
corresponding to a frequency resolution of 24.4 kHz (127 channels).  The
noise in line free channels was $\sim$ 27 mJy/beam. In this case we
estimate the calibration accuracy to be no better than 20\%.

We re-reduced some snapshot CH$_3$OH 7$_0$ $\to$ 6$_1$ A$^+$
observations from 17 September 2001 at 44 GHz, which show a class I
methanol maser in the vicinity of IRS\,9 \citep{Schwartz02}. The
synthesized beam was 1\ptsec8 $\times$ 0\ptsec76 (p.a. $+$53\degr{}).  The
velocity resolution for these observations was 0.4 km~s$^{-1}$ and the
velocity coverage was 30 km~s$^{-1}$. The RMS noise was 40 -- 55
mJy~beam$^{-1}$ in line free channels.

\subsection{JCMT Archive Data}

We retrieved the CO \jtra21 data obtained with the James Clerk Maxwell
Telescope (JCMT) on Mauna Kea, Hawaii, on 1997 June 2 \& 3 from the JCMT
archive at CADC\footnote{Guest User, Canadian Astronomy Data Center,
which is operated by the Dominion Astrophysical Observatory for the
National Research Council of Canada's Herzberg Institute of
Astrophysics}; see \citet{Davis98} for a more detailed description. This
observation was a an on--the--fly map of the whole NGC\,7538 molecular
cloud scanned at a position angle of $-$45\degr\ sampling every
7\arcsec\ in the scan direction and stepping 10\arcsec\ between scans.
The integration time per point was 12 seconds. Due to the sparse
sampling the beam is somewhat broadened in the scan direction. We
estimate the  HPBW to be $\sim$ 22\arcsec\ $\times$ 21\arcsec. The
velocity resolution was 0.49 km~s$^{-1}$ and the velocity coverage was
650 km~s$^{-1}$ (500 MHz bandwidth). The data were reduced with SPECX
and then exported to MIRIAD for further analysis. For the majority of
the scans we only subtracted a linear baseline, but some of the scans
taken towards the end of the run had poorer baseline quality. For these
scans we used a fifth order polynomial, which resulted in flat
baselines. The final map has an RMS  of 1.5 K~km~s$^{-1}$ in T$_{mb}$.

We also found some long integration H$_2$CO data of good quality taken
on 1995 November 4. The H$_2$CO  $3_{2,2} - 2_{2,1}$ transition at
218.47564 GHz and the $3_{0,3} - 2_{0,2}$ transition at 218.22219 GHz
were observed simultaneously in a 500 MHz wide band, giving a velocity
resolution of $\sim$ 0.48  km~s$^{-1}$. The ratio of these H$_2$CO 
transitions is a good temperature probe for cold or moderately hot
gas \citep{Mangum93}. The integration time towards IRS\,9 was 1980 sec,
resulting in an RMS of 0.11~K  in T$_{mb}$. The same data set also had an
even longer integration, 3,000 sec, at the position ($+$12\arcsec, $-$35\arcsec{})
relative to IRS\,9. The noise level of this spectrum is 0.08 K.

\section{Continuum results}
\label{sect-cont}

The deep  3.6 cm VLA observations with an angular resolution of
1\ptsec21 $\times$ 0\ptsec47 show a clear detection of a faint,
marginally resolved source centered on IRS\,9 (Table \ref{tbl-1}). There
are no other sources within 100\arcsec\ of IRS\,9.  At 6~cm the
free--free emission from IRS\,9 is only a 3--$\sigma$ detection.
However, since the 6~cm emission agrees well in position with the 3.6~cm
emission, we consider it to be realiable. The flux density at 6~cm is
comparable to the 3.6~cm data, suggesting a flat spectral index. Such a
flat spectral index is also found for the protostar NGC\,7538\,S
(Sandell et al., in preparation) and most likely due to free-free
emission from a well-collimated ionized wind \citep{Reynolds86}.

At 3.4~mm we find the continuum emission from IRS\,9 to be a
point-source surrounded by weak extended emission. The continuum peak
agrees extremely well with our VLA position (Fig. \ref{fig-cont}, Table
\ref{tbl-1}). Since the free-free emission is very weak and has a
spectral index $\sim$ 0, essentially all the emission at 3.4~mm must
arise from thermal dust emission. The observed size of the dust emission
is about twice as large as  \citet{Sandell04} determined from single
dish observations at 1.3~mm, 850 and 450 $\mu$m, suggesting that most of
the dust emission at 3.4 mm originates from the dense cloud core in
which IRS\,9 is embedded. We note that our results disagree with the
results reported by \citet{Tak00}, who found the dust emission to be
unresolved in high spatial resolution ($\sim$ 2\arcsec\ beam) OVRO
observations at 107 and 115 GHz (Table \ref{tbl-1}). Their flux
densities, 43 and 95 mJy, at 107 and 115 GHz, respectively, are at least
a factor of two higher than the current BIMA observations. If we
extrapolate from the results quoted by \citeauthor{Sandell04}, who
derived a dust emissivity index, $\beta$ $\sim$ 2, the single dish
observations predict a integrated flux density of $\sim$ 18 mJy,  much
less than our integrated flux density of 85 mJy at 3.4~mm (Table
\ref{tbl-1}). The $\beta$-index derived by \citet{Sandell04} is
therefore likely to be overestimated. Most protostellar sources have an
apparent $\beta$-index of 1 to 1.5 \citep{Shirley00}, and a
$\beta$-index of $\sim$2 is not expected. In the following we assume
that the $\beta$-index is 1.5.

If we assume that the dust emission is optically thin and that the bulk
of the gas is at an average temperature, T$_d$, the total mass of gas
and dust, M,  can be expressed as  ${\rm M = S_\nu D^2/(\kappa_\nu
B_\nu(T_d))}$.  Here  ${\rm B_\nu(T_d)}$ is the Planck function,
$\kappa_\nu$ is the dust mass opacity,  $D$ is the distance, and S$_\nu$
is the integrated flux density at the frequency $\nu$.  For $\beta$ =
1.5 and a gas to dust ratio of 100 we get $\kappa_{87} $= 0.002 $cm^2
g^{-1}$ \citep{Hildebrand83}.  A dust temperature,T$_d$ of $\sim$ 40~K
seems to be appropriate, based both on the results by \citet{Sandell04}
and the average gas temperature in the vicinity of IRS\,9, which is
$\sim$ 30~K \citep{Hasegawa95}.  With these assumptions we find a total
mass from the 3.4~mm dust emission of $\sim$ 180\Msun. If the
$\beta$-index was as high as 2, the mass would be about four times
larger. Since we find that IRS\,9 drives a well-collimated molecular
outflow (Section \ref{sect-flow}), we assume that IRS\,9 is surrounded
by an accretion disk. In this case the disk is most likely represented
by the unresolved point-source emission at 3.4~mm ($\sim$ 8 mJy), and
not by the extended emission, which originates from the surrounding
dense cloud envelope. Our continuum observations therefore predict a
disk--mass of $\sim$20\Msun.  This estimate is quite uncertain, both due
to the uncertainty in the flux density assigned to the disk, and because
of the limited knowledge of the dust properties and dust temperature of
the proposed disk.

\section{Molecular Cloud Cores - H$^{13}$CN and SO emission}
\label{sect-cores}

H$^{13}$CN \jtra10 and SO 2$_2\to1_1$ are good tracers of high density
gas, i.e. densities $\geq$ 10$^5$ cm$^{-3}$. H$^{13}$CN is  a chemically
robust molecule, i.e. its abundance is not strongly affected by shocks
or temperature conditions. The emission should be optically thin and
therefore provide an accurate image of the column density distribution
of a dense molecular core as long as the gas temperature is fairly
constant in the core. SO, although quite abundant in molecular clouds,
often shows a different column density distribution relative to other
high density tracers, presumably due to chemical effects
\citep{Swade89}. The SO abundance is also enhanced in molecular outflows
\citep{Plambeck82,Chernin94}, and can be significantly enhanced even in
low-velocity shocks \citep{Mitchell84}.

HCO$^+$ traces somewhat lower gas densities, $\sim$ 10$^4$ cm$^{-3}$,
than H$^{13}$CN and is expected to be  optically thick.
It is therefore not a reliable column density tracer, at least not for the very
dense molecular cloud cores that surround young high-mass stars.
Furthermore the HCO$^+$ abundance is often enhanced in high-velocity
outflows \citep{Mitchell91}, and the high velocity gas may completely
dominate, as is the case towards IRS\,9, see Section \ref{sect-flow}.

NH$_2$D 1$_{11} \to1_{01}$ traces primarily cold, pristine molecular gas
\citep{Shah01}. Sandell et al. (2005, in preparation) have detected
NH$_2$D in the NGC\,7538 molecular cloud  towards the high-mass
protostar NGC\,7538\,S and in the dust filaments east and west of
NGC\,7538\,S. No NH$_2$D emission is detected in the molecular cloud
core near IRS\,9;  the IRS\,9 core is therefore not likely to have a
significant amount of cold, pristine gas.

Our observations show several regions of dense gas in the IRS\,9 field
(Fig. \ref{fig-h13cn-so}). There is an extended cloud core SE of
IRS\,9. The star itself is embedded in a ``compact'' cloud core almost
detached from the larger cloud. To the west, near the edge of our
primary beam, there is a region of very strong, narrow SO emission. High
resolution images of NH$_3$ \citep{Zheng01} show a very similar
distribution, except that NH$_3$ is largely absent towards IRS\,9,
whereas we detect the strongest  H$^{13}$CN emission towards IRS\,9.

\subsection{The IRS\,9 Cloud Core}

The  H$^{13}$CN and SO emission peak within $\sim$ 1\arcsec\   of
IRS\,9. The SO emission appears more compact, and shows a  fainter
secondary peak  $\sim$ 5\arcsec\ W of IRS\,9  (Fig \ref{fig-h13cn-so}).
The size of the H$^{13}$CN emission, excluding high velocity gas, is
$\sim$ 18\arcsec\ $\times$ 8\arcsec\ extending EW, i.e. similar to the
extent of the dust emission at 3.4~mm (c.f. Section \ref{sect-cont}).
The radial velocity determined from the main hyperfine transition of
H$^{13}$CN (F =$2\to1$) , $\sim$ $-$57 km~s$^{-1}$, agrees well with
that determined from observations of a number of transitions of CS and
H$_2$CO  \citep{Tak00}. We resolve the hyperfine structure in H$^{13}$CN
 and in addition to the F = $2\to1$ transition we also observe the upper
hyperfine transition, F =  $1\to1$ (Fig.  \ref{fig-irs9-spectra}), which
for optically thin emission has a line intensity of 0.6 times the F =
$2\to1$ line. The F = $0\to1$ hyperfine component may also be present,
but is generally weak ($\sim$ 0.2 times the intensity of the F = $2\to1$
transition for optically thin emission).  Since the  F =  $1\to1$
hyperfine transition is only separated by $+$4.9 km~s$^{-1}$ from the F
= $2\to1$ transition, the two lines often appear partially blended. We
observe no obvious velocity gradient in either H$^{13}$CN or SO. Both 
H$^{13}$CN and SO show high velocity wings (Fig.
\ref{fig-irs9-spectra}), confirming that the extreme high velocity
outflow driven by IRS\,9 has very high densities, at least at low and
moderate velocities. The high velocity emission completely dominates in
HCO$^+$ (Fig. \ref{fig-irs9-spectra}), which will be discussed in more
detail in Section \ref{sect-extreme-flow}.

The single dish JCMT H$_2$CO data (Fig. \ref{fig-h2co}) give
T$_{mb}$($3_{0,3} - 2_{0,2}$)/T$_{mb}$($3_{2,2} - 2_{2,1}$) = 3.5 $\pm$
0.3, integrated over the central 3 km~s$^{-1}$, suggesting a gas
temperature $\ge$ 60~K \citep{Mangum93}. However, \citet{Hasegawa95}
derive a lower temperature of 32 $\pm$ 8 K from analysis of the \jtra32
and \jtra43 transitions of HCO$^+$ and H$^{13}$CO$^+$, while
\citet{Mitchell90}, using absorption lines of the fundamental
vibrational band of $^{13}$CO at 4.7 $\mu$m, derived 26 K. The latter
two studies therefore suggest that the temperature of the dense gas
surrounding IRS\,9 is $\sim$ 30~K, while the H$_2$CO data suggest a
substantially higher temperature. The other position observed in
H$_2$CO, at ($+$12\arcsec,$-$35\arcsec\ ) in the southern cloud core
gives T$_{mb}$($3_{0,3} - 2_{0,2}$)/T$_{mb}$($3_{2,2} - 2_{2,1}$) = 7.0
$\pm$ 0.4, corresponding to a gas temperature of $\sim$ 30 -- 35 K, in
good agreement with the temperature derived by  \citet{Hasegawa95}  from
HCO$^+$ data at the same position. However, since both H$_2$CO
transitions show clear line wings towards IRS\,9,  the  emission from
the hot outflow  at near-cloud velocities add a substantial contribution
to the line integral, causing an incorrect line ratio for the core
emission. If we fit the H$_2$CO spectra with a two-component Gaussian,
one for the cloud emission and one for the broad wing emission, we find
that a ratio of  $\sim$8 for the cloud core, corresponding to an average
gas temperature of 30 K. In the follwing we adopt 30 K as the
excitation temperature for H$^{13}$CN, resulting in a column density
towards IRS\,9 of $\sim$ 3.6~10$^{13}$ cm$^{-2}$.  If we assume an
abundance ratio for HCN similar to the  OMC-1 extended ridge
\citep{Blake87}, i.e. [HCN]/[H$_2$] = 5 10$^{-9}$, and an isotope ratio,
[$^{12}$C]/[$^{13}$C] = 85 \citep{Wilson94}, we find a total H$_2$
column density of 6.1~10$^{23}$ cm$^{-2}$ towards IRS\,9.  If we assume
that the relationship between color excess and total hydrogen column
density \citep{Bohlin87} is valid also for dense molecular gas, the
observed column density suggests that IRS\,9 is obscured by a
visual extinction of more than 300$^m$, yet \citet{Tielens91} derived a
visual extinction of only $\sim$ 75$^m$ from observations of the CO
absorption band at 4.67 $\mu$m. These two estimates are not necessarily
discrepant, because as we will show in Section \ref{sect-extreme-flow},
IRS\,9 drives a well collimated high velocity outflow observed almost
pole-on, clearing away enough of the foreground cloud to allow the star
to be detectable in the near IR.

If we integrate the H$^{13}$CN emission over the IRS\,9 cloud core we   
derive  a total mass  $\sim$ 350 \Msun. This mass estimate is in good
agreement with the value based on our continuum observations, 180
\Msun\ (Section \ref{sect-cont}), especially considering that the the
abundance ratio of [H$^{13}$CN/H$_2$] is uncertain by at least a factor
of two or three. Our observations therefore show that  IRS\,9 is still
surrounded by a very dense cloud core  with  a mass of $\geq$ 100
\Msun.

\subsection{The cloud core SE of IRS\,9}
\label{sect-protostar}

The molecular cloud core SE of IRS\,9 dominates the emission in
H$^{13}$CN  and SO. The core has a diameter  of $\sim$ 40\arcsec\  and
shows several emission peaks. H$^{13}$CN  shows a ridge of emission in
p.a. $\sim$ 25 \degr\ , with a strong peak at an offset $\sim$
(6\arcsec, $-$22\arcsec{}) from IRS\,9 (Fig \ref{fig-h13cn-so}).
Another extended emission peak  is observed at the SE boundary of the
H$^{13}$CN cloud and is also distinctly visible on the 450 $\mu$m SCUBA
image (Fig. \ref{fig-h13cn+450}). It is barely detected in SO, which
otherwise exhibits a similar morphology (Fig. \ref{fig-h13cn-so}).
Both H$^{13}$CN and SO show red- and
blue-shifted wings towards the dominant emission peak at (6\arcsec,
$-$22\arcsec{}) ( Fig \ref{fig-irs9-spectra}), while HCO$^+$ shows deep
red-shifted self-absorption. Even H$^{13}$CN is strongly affected by
self-absorption. Strong red-shifted self-absorption in HCO$^+$ is
usually interpreted as evidence for infall or accretion towards a
central protostellar object. Since we also observe a bipolar outflow
originating from this position (Section \ref{sect-flow}),  there must be
a young protostar deeply embedded in this cloud core. A Gaussian fit to
the H$^{13}$CN peak gives a size of 14\arcsec\ $\times$ 6\arcsec\  (p.a.
$\sim$ 20\degr{}). If we assume the same gas temperature and H$^{13}$CN
abundance as used for the cloud core surrounding IRS\,9, a mass of $>$
250 \Msun\ for the compact cloud core surrounding the protostar is
suggested. This is a lower limit, since we have made no correction for
the self-absorption, which virtually blocks all emission at the cloud
velocity (Fig \ref{fig-protostar}). The average gas density of this
protostellar core is therefore $>$ 6~10$^6$ cm$^{-3}$. The large cloud
core, in which this protostellar core is embedded, has a mass of $>$ 900
\Msun, providing a large mass reservoir for additional star formation.

 We have have examined the H$^{13}$CN and SO data to see if there is any
 evidence for a rotating disk, which would be expected around such a
 young, heavily accreting protostar. However, based on the position
 velocity plot in Fig. \ref{fig-protostar}, the dense surrounding  cloud
 core effectively obscures the protostar; it is thus impossible to
 observe the central protostar in these molecular transitions. An
 improved ``disk probe" is therefore required, which is not obscured by
 the cold cloud envelope or the outflow, which may mask any kinematics
 of the disk.

 Such a disk maybe observable in CH$_3$CN or in higher rotational
 transitions of H$^{13}$CN or DCN. However, since one can see high
 velocity emission from the dense outflow in H$^{13}$CN \jtra10, and
 since the outflow appears to be hot (Sect. \ref{sect-temp}), the
 outflow is likely to dominate, making H$^{13}$CN a poor choice.
 DCN is usually not seen in outflows, but appears to be a
 good disk tracer (Sandell et al. 2005, in prep.) and may therefore be a better
 choice than H$^{13}$CN.

\subsection{The sub-mm source IRS\,9\,S}

H$^{13}$CN  and SO avoid  IRS\,9\,S, the southern sub-mm source, which
\citet{Sandell04} suggest might host another protostar. Both H$^{13}$CN
and SO seem to form almost a semi-shell around the eastern boundary of
this sub-mm source (Fig. \ref{fig-h13cn+450}). The absence of molecular
emission from IRS\,9\,S is puzzling. One possibility is that this is an
extremely dense, cold protostellar core, in which most molecules are
frozen onto interstellar dust. However, if this is the case, we would
expect to see NH$_2$D emission, which is absent. If IRS\,9\,S  were a
cold source, strong emission strong source at 1.3mm and 850 $\mu$m is
expected. However,  \citet{Sandell04} found it to be faint at these
wavelengths. The other possibility is that IRS\,9\,S is a hot source. We
note that \citet{Zheng01} detected NH$_3$ towards IRS\,9\,S, and remark
that the NH$_3$ emission shows larger velocity dispersion and higher
optical depth than the emission in the vicinity of IRS\,9. A plot of the
line ratios of the (J,K) inversion transitions (2,2) and (1,1) presented
by \citet{Zheng01} indicates that the gas is hotter towards IRS\,9\,S. A
dense hot  (T$_R$ $\ge$ 80 K) core is a more likely explanation. Yet the source of
energy for such a hot core is unknown, since we detect no free-free
emission at 3.6~cm to a limit of 60 $\mu$Jy~beam$^{-1}$, nor is 
any outflow or maser emission present in the core.

\section{High velocity outflows and masers}
\label{sect-flow}

The BIMA HCO$^+$ \jtra10 spectrum towards IRS\,9 shows broad line wings
(Fig. \ref{fig-irs9-spectra}); we have not even observed the full extent
of the high velocity emission in our 25 MHz band, corresponding to a
velocity range of 90 km~s$^{-1}$ ($-$94 to $-$11~km~s$^{-1}$). The broad
line wings are no surprise, since IRS\,9 is known to power an extreme
high velocity CO outflow \citep{Mitchell91}, which is also very
prominent in HCO$^+$ \citep{Hasegawa95}. The extreme high velocity
emission is only observed in the vicinity of IRS\,9, although there is
widespread high velocity gas in the IRS\,9 molecular core
\citep{Kameya90,Mitchell91} and several regions of shocked H$_2$
emission including a well collimated jet  \citep{Davis98}.
\citet{Kameya89} noticed that the high velocity gas $\sim$ 30\arcsec\ SE
of IRS\,9 showed both blue- and red-shifted emission, suggesting that
this gas could be part of a separate outflow. \citet{Davis98} also
suggested that multiple outflows exist in the IRS\,9 region.

In the following we will use our high spatial resolution HCO$^+$ \jtra10
BIMA images together with images of shocked 2.12 $\mu$m H$_2$ and single
dish CO  \jtra21 images from \citet{Davis98}, to show that there are at
least three separate high velocity outflows in the IRS\,9 region, all
partially overlapping with each other. The occurrence of H$_2$O, 1720
MHz OH, and class I CH$_3$OH masers, which are excited in these
outflows, provide additional information in identifying the various
outflows. The H$_2$O masers, which may be excited in the accretion disks
powering the outflows, provide further evidence on the location of the
stars driving the outflows.

To obtain a better overview of the distribution of the high velocity
gas, we use four sets of images of  HCO and CO (Fig \ref{fig-outflows}),
where we partitioned the blue- and red-shifted  high velocity gas into
four velocity regimes: {\it i)} low velocities, {\it ii)} intermediate
velocities, {\it ii)} high velocities, and {\it iv)}extreme velocities.
The first three sets are all integrated over  6.5 km s$^{-1}$ wide
velocity intervals for both HCO$^+$ and and CO. We begin at velocities
of $\pm$ 5 km~s$^{-1}$ from the cloud core to avoid emission from
turbulent gas in the ambient cloud.  For  the extreme high velocity
emission we begin at velocities $\pm$ 25 km s$^{-1}$ from the cloud
velocity and integrate to the highest velocities detected. These contour
images of blue- and red-shifted emission are shown in Fig.
\ref{fig-outflows}.

Using the HCO$^+$ (and  CO) images we identify three outflows: {\it i)}
The extreme high velocity HCO$^+$ emission shows a compact bipolar
outflow centered on IRS\,9 (the IRS\,9 extreme high velocity outflow),
{\it ii)} At  low and intermediate velocities, HCO$^+$ shows a highly
collimated  blue-shifted outflow well aligned with the H$_2$ jet at a
p.a. of  155\degr\ with no red-shifted counterpart (the H$_2$ jet), and
{\it iii)} a poorly collimated blue-shifted outflow with a compact
red-shifted counterflow at p.a. $\sim$ 28 \degr\ (the protostellar
outflow). The last outflow intersects with the H$_2$ jet $\sim$
24\arcsec\ south of IRS\,9. In the following we discuss each outflow in
more detail, adding information from the CO images, the maser
observations and images of high density gas based on H$^{13}$CN and SO,
which were discussed in Section \ref{sect-cores}.

\subsection{The IRS\,9 extreme high velocity outflow}
\label{sect-extreme-flow}

The HCO$^+$ data show that IRS\,9 drives a bipolar extreme high velocity
outflow approximately oriented EW. The center of symmetry  is within
1\arcsec\ of IRS\,9 (Fig. \ref{fig-outflows}). The outflow is very
compact and highly collimated; the extent of the outflow lobes are
$\sim$ 14\arcsec. The two H$_2$O maser spots  (Table \ref{tbl-2})
detected by \citet{Kameya90} are centered close to the VLA continuum
source. Both are blue-shifted relative to the cloud velocity and are
probably excited in the outflow close to the stellar disk. We also
observe two CH$_3$OH masers, one close to the star and one in the
blue-shifted outflow lobe. Both have velocities close the the systemic
velocity. We also detect a slightly red-shifted 1720 MHz OH maser in the
red-shifted outflow lobe. \citeauthor{Davis98} detected two bright H$_2$
knots west of IRS\,9. Both lie in the blue-shifted outflow lobe at
$\sim$ ($-$9\ptsec1, $-$3\ptsec6) and ($-$16\ptsec6, $-$2\ptsec5)
relative to IRS\,9, suggesting that they are excited by the  high
velocity gas in the blue-shifted outflow. Since the outflow lobes of the
IRS\,9 outflow are largely overlapping, the HCO$^+$ observations suggest
that the extreme high velocity  outflow is observed nearly pole on. If
we assume that the ellipticity of the outflow lobes is caused by
projection, the outflow is inclined by $\sim$ 20\degr\ to the line of
sight. Observing the outflow pole on explains why the IRS\,9 outflow
shows such extreme high velocities. \citet{Mitchell91} found that the
red-shifted outflow has an outflow velocity of 110 km~s$^{-1}$ in CO
\jtra32, while the blue-shifted outflow wing was much less prominent and
has a velocity span of only 40  km~s$^{-1}$. In the single dish CO
\jtra21 images  (Fig. \ref{fig-outflows}), the outflow is severely
blended with the two other outflows and the bipolar nature is not at all
evident. At the position of IRS\,9, however, the CO high velocity wings
extend from $\sim$ $-$ 80 to $+$20 km~s$^{-1}$ in V$_{lsr}$, with a
strong red-shifted wing and a much fainter  blue-shifted wing. In
HCO$^+$ the blue- and red-shifted outflow wings are largely symmetrical
(Fig. \ref{fig-irs9-spectra}). The approximate symmetry of the two
outflows lobes is clear in the position velocity plot taken along the
symmetry axis of the outflow (Fig. \ref{fig-cut80}).

Since our observations show that the outflow is much more compact than
previously believed, it is also much younger. If we use the observed
maximum velocity of 110 km~s$^{-1}$ \citep{Mitchell91} and assume
20\arcsec\  as the total extent of the outflow ($\sim$ 0.8 pc), we find
a dynamical timescale of $\sim$ 6600 yr for the red outflow lobe. The
large mass outflow and youth of the IRS\,9 outflow suggests that IRS\,9
is still in an active accretion phase. Even though the outflow is
observed almost pole on, and the outflow therefore largely masks any
infall signature, a narrow absorption in the redshifted outflow lobe is
detected (Fig. \ref{fig-irs9-spectra}), most likely arising from cold
infalling gas (see Section \ref{sect-accretion}).

\subsection{The H$_2$ jet}
\label{sect-h2-jet}

The narrowband 2.12 $\mu$m H$_2$ imaging by \citet{Davis98} shows a
highly collimated H$_2$ jet terminating $\sim$ 60\arcsec\ S of IRS\,9.
As we have already discussed, this jet is not powered by IRS\,9, because
the high spatial resolution HCO$^+$ imaging shows that IRS\,9 drives a
compact nearly pole-on bipolar outflow. At intermediate blue-shifted
velocities HCO$^+$ shows an extremely well collimated outflow at a p.a.
$\sim$ 155\degr, which coincides with the H$_2$ jet (Fig.
\ref{fig-outflows}). This outflow is also observed in CO, where it
dominates the high velocity emission at all blue-shifted velocities
(Fig. \ref{fig-outflows}). With a  22\arcsec\ resolution, the outflow
blends in with the IRS\,9 outflow, giving the appearance that it is
powered by IRS\,9. The CO images also show a prominent red-shifted
counter jet NW of IRS\,9 extending to $\sim$ 50\arcsec\ N of IRS\,9.
This red-shifted outflow lobe has no obvious counterpart in HCO$^+$. At
low and intermediate velocities, faint red-shifted HCO$^+$ emission
coinciding with the cluster of H$_2$ knots $\sim$ 50\arcsec\ N of IRS\,9
and at ($\sim$ $-$30\arcsec, $+$20\arcsec{}) (Fig.
\ref{fig-h2+hco+h13cn}) is observed. Both of these compact red-shifted
HCO$^+$ knots lie in the outskirts of the red-shifted CO lobe and
coincide with regions of high gas density. At both positions where we
observe HCO$^+$ high velocity emission N of IRS\,9, we also detect
H$^{13}$CN emission.

At low velocities a compact red-shifted emission peak slightly north of
the methanol maser IRS\,9-CH$_3$OH(3) (Fig. \ref{fig-outflows}) is
obvious. This red-shifted emission peak is close to the protostellar
source at (6\arcsec, $-$22\arcsec{}). This coincidence is seen more
clearly in Fig. \ref{fig-h2+hco+h13cn}, where we overlaid the HCO$^+$
emission integrated over low and intermediate velocities on the  map of
integrated H$^{13}$CN emission. The CO image, however, shows that this
red-shifted HCO$^+$ emission peak coincides with a red-shifted CO
outflow extending to the SW rather than to the N. The red-shifted
HCO$^+$ emission therefore appears to be part of the protostellar
outflow. The source of the blue-shifted HCO$^+$ may reside in the
protostellar core. However, more likely the source of the outflow lies
in the IRS\,9 cloud core, since we observe high velocity blue-shifted
HCO$^+$ emission between the protostellar core and  IRS\,9. The 2.12
$\mu$m H$_2$ image shows a nebulous IR source at  an offset of
($-$2\ptsec2, $-$1\arcsec{}) from IRS\,9, labeled as peak A1 by
\citet{Tamura91}, who intepreted it as a peak in the reflection nebula.
We believe it may be another embedded young star and therefore a
potential exciting source for the HCO$^+$ jet.

This outflow looks very similar to the IRS\,9 outflow, if we account for
the difference in inlination. Both outflows are very highly collimated
and jet-like and similar in size.  The CO map shows a maximum velocity
of  $\sim$ 50 km~s$^{-1}$ for the blue-shifted H$_2$ jet.  This velocity
may be a lower limit, since the CO image does not cover the full extent
of the blue outflow lobe; a position velocity plot of CO along the
outflow shows that the outflow is accelerating, i.e. the highest outflow
velocities are observed at the tip of the outflow. The same behavior is
seen in HCO$^+$, but the maximum velocity in HCO$^+$ is lower by $\sim$
25  km~s$^{-1}$. The velocities are somewhat lower in the red-shifted
outflow, where CO  shows a maximum velocity of $\sim$ 37 km~s$^{-1}$.
The outflow is quite extended, $\sim$  50\arcsec\ -- 70\arcsec,
depending on whether we assume that the driving source lies in the
protostellar core or in the IRS\,9 core.

\subsection{The protostellar outflow}
\label{sect-protostellar_outflow}

At extreme velocities (Fig. \ref{fig-outflows})  faint blue-shifted
HCO$^+$ emission extending from the southern H$^{13}$CN peak to the NE
(p.a. $\sim$ 28\degr{}) is observed.  The maximum velocity of the
blue-shifted HCO$^+$ outflow is $\sim$ 28 km~s$^{-1}$. At lower
velocities the outflow lobe merges into the outflow tracing the H$_2$
jet and appears less collimated. At low velocities HCO$^+$ shows a more
compact lobe of red-shifted emission, which is well aligned with the
symmetry axis of the blue-shifted emission. The intercept between red-
and blue-shifted emission is roughly at the position of the southern
H$^{13}$CN peak (Fig. \ref{fig-h2+hco+h13cn}). However, the red-shifted
HCO$^+$ emission is also approximately on the symmetry axis of the H$_2$
jet, and could therefore also be associated with the  H$_2$ jet.

In the low spatial resolution CO maps the blue-shifted emission is
completely blended with the outflow associated with the H$_2$ jet. The
only appearance of this rather compact blue-shifted HCO$^+$ outflow in
CO is that the blue CO wing covers a larger velocity range ($\sim$ 35
km~s$^{-1}$) at the position of the protostar and the blue outflow lobe
widens up and curves to the N just S of IRS\,9. The red-shifted counter
flow, which is very compact in HCO$^+$, is clearly visible in CO  at
moderate and high velocities. At higher velocities the red-shifted
emission lobe appears to be shifted to the N (Fig. \ref{fig-outflows}).
It is therefore possible that what we identified as the symmetry axis of
the blue-shifted outflow is the northern wall of the outflow. An outflow
with a p.a. of  60\degr{}, and a poorly collimated blue outflow lobe
with an opening angle of 50\degr\ -- 60\degr{} is also plausible. Such an
outflow geometry could explain why we observe high velocity gas at the
SW boundary of the H$^{13}$CN core, which in this scenario would
represent the southern outflow wall of the outflow.

As we already noted in Section \ref{sect-protostar}, the southern
H$^{13}$CN peak at (6\arcsec,$-$22\arcsec\ ) shows high velocity wings in
both H$^{13}$CN and SO (Fig. \ref{fig-irs9-spectra}), suggesting that
there is a YSO embedded in the core driving a high velocity outflow.
This YSO appears to be a protostar, since HCO$^+$ shows deep red-shifted
self absorption, presenting strong evidence for infall (Section
\ref{sect-accretion}). The emission from dense high velocity gas traced
by H$^{13}$CN places the protostellar source right between the red- and
the blue-shifted outflow lobe (Fig. \ref{fig-h13cn-high}), confirming
that the protostellar source drives this rather compact outflow, which
in the blue-shifted outflow lobe only extends $\sim$ 25\arcsec\ --
30\arcsec\ from the driving source  terminating near the eastern
red-shifted outflow lobe from IRS\,9.

The protostellar outflow shows deceleration, i.e. the highest outflow
velocities are seen towards the center of the outflow. In this respect
it looks very similar to the very young compact outflow driven by
NGC\,7538\,S \citep{Sandell03}. If molecular outflows from high-mass
protostars are driven by jets, one would expect to see deceleration
when the outflow is very young and has not yet had time to accelerate
the gas in the surrounding dense cloud core. This outflow may therefore
be even younger than the IRS\,9 outflow, even though its dynamical
timescale is about the same or larger than the IRS\,9 outflow (Table ~\ref{tbl-3}).
Both outflows are certainly very young.

\subsection{Unexplained high velocity features}

At low red-shifted velocities the CO map shows a tongue of emission
extending to $\sim$ 50\arcsec\ west of IRS\,9 (Fig. \ref{fig-outflows}).
This velocity feature is also seen in the CO \jtra32 map by
\citet{Mitchell91}. Although we can plausibly explain the shocked
H$_2$ knots north of IRS\,9 by assuming the red-shifted outflow is
expanding more freely through the low density gas north of IRS\,9, it
cannot explain why the outflow would flow backwards. It is therefore
likely that there is another outflow west or northwest of IRS\,9. This
outflow terminates (or originates) close to the HH object reported by
\citep{Campbell88}, and would therefore provide the energy to excite
this HH object. Our HCO$^+$ map also shows low velocity red- and
blue-shifted emission  at $\sim$ ($-$33\arcsec,$-$30\arcsec{}) in the
southwestern part of the reflection nebula, which has the appearance of
a compact bipolar outflow (Fig. \ref{fig-h2+hco+h13cn}). This outflow is
not seen in CO, which makes it somewhat dubious. However, if the outflow
is as compact and faint as indicated by our HCO$^+$ data, it would be
difficult to see in the CO map, which does not go very deep.

\subsection{Outflow characteristics}

We derive the physical characteristics of the outflows following the
recommendations and formulae presented in the careful study of outflow
characteristics by \citet{Cabrit92}. Since we have images in both
HCO$^+$ and CO we derive the outflow parameters by integrating over the
same velocity intervals that we used for identifying the outflows, see
Section \ref{sect-flow}.   For correcting the outflow parameters for
inclination, we assume a mean inclination of 57.3\degr, see e.g.
\citet{Bontemps96}, except for the IRS\,9 outflow, where we use 20\degr,
as determined from the outflow morphology (Sect. \ref
{sect-extreme-flow}). However, most of our information about the
outflows comes from HCO$^+$ and since the HCO$^+$ abundance can be
considerably enhanced in outflows \citep{Garden92,Hasegawa95}, this
could lead to large uncertainties in determining the mass of the high
velocity gas. The IRS\,9 outflow is also known to be very hot
\citep{Mitchell91} and since the column density of a molecule is
approximately linearly dependent on the gas temperature, the mass of an
outflow can be seriously underestimated if  the outflow temperature is
assumed to be the same temperature as the surrounding molecular cloud;
an assumption often made in outflow studies
\citep{Shepherd96,Beuther02a}.  Below we discuss in more detail how we
determined the temperature and the mass of the outflows, which have been
used to derive most of the outflow parameters given in Table
\ref{tbl-3}.

\subsubsection{Temperatures of the  outflows}
\label{sect-temp}

\citet{Mitchell91} used their CO \jtra32 and \jtra21 observations to
deduce the gas temperature of the IRS\,9  high velocity outflow. They
found the red-shifted gas to be  very hot, $\sim $ 200~K, while they
deduced a much lower temperature for the blue-shifted gas, $\sim$ 40 K.
A later study \citep{Hasegawa95}, using several rotational transitions
of HCO$^+$, gave similar results.  \citet{Mitchell90} observed the
fundamental vibrational band of CO and $^{13}$CO at 4.6 $\mu$m  in
absorption against IRS\,9 and found a temperature of $\sim$ 180 K, which
is in good agreement with what \citet{Mitchell91} found for the
red-shifted outflow using the rotational CO emission lines. However,
since the absorption studies must  primarily sample the blue-shifted
gas, there is an inconsistency between the hot temperature deduced from
the absorption studies and the cold temperature deduced from emission
lines. The apparent weakness of the blue-shifted CO wing compared to the
red-shifted  wing is easily explained if we assume that the hot gas in
the blue outflow is self-absorbed against colder high velocity gas, see
e.g. \citet{Beltran04}. This explains not only the weakness of the
blue-shifted outflow relative to the red-shifted outflow, but also
why the  emission lines produce an incorrect temperature for
the blue-shifted outflow. The H$_2$CO  $3_{2,2} - 2_{2,1}$ and $3_{0,3}
- 2_{0,2}$ transitions only show weak line wings towards IRS\,9 (Fig.
\ref{fig-h2co}), but the wing emission is about equally strong in both
transitions and very symmetric. A two component Gaussian fit to both
spectra give the same radial velocity for both the cloud core and
plateau (wing emission) resulting in a line ratio T$_{peak}$($3_{0,3} -
2_{0,2}$)/T$_{peak}$($3_{2,2} - 2_{2,1}$) = 1.9 for the high velocity
emission, strongly supporting the earlier conclusion that the gas is hot
in both outflow lobes. If the blue-shifted outflow was as cold as 40~K,
we would not see any blue-shifted wing in  the $3_{2,2} - 2_{2,1}$
transition. For estimating outflow masses we will therefore adopt an
excitation temperature of 180~K for the red- and the blue-shifted high
velocity emission.

For the other two outflows we have far less information, but both appear
colder than the IRS\,9 outflow. \citet{Hasegawa95} derived a temperature
of 40~K for the position (+10\arcsec, $-$35\arcsec{}), which is well
centered on the blue-shifted jet, although  within their 15\arcsec-beam,
they also include some emission from the protostellar outflow. The
spectra of the $3_{2,2} - 2_{2,1}$ and $3_{0,3} - 2_{0,2}$ transitions
of H$_2$CO at the same position (Fig. \ref{fig-h2co})  show strong
blue-shifted wings in both transitions and the ratio of the wing
emission integrated from $-$72 -- $-$60 km~s$^{-1}$, i.e.
T$_{wing}$($3_{0,3} - 2_{0,2}$)/T$_{wing}$($3_{2,2} - 2_{2,1}$) = 3.13
$\pm$ 0.06, suggesting a temperature $\ge$ 80~K \citep{Mangum93},
seemingly in conflict with the HCO$^+$ results. However, in this case
the H$_2$CO HPBW is larger, 22\arcsec, and the high velocity emission
seen in H$_2$CO must originate from the protostellar outflow, not the
H$_2$ jet, because only the protostellar outflow has outflow densities
which are high enough to excite H$_2$CO in the outflow. For the H$_2$
jet we therefore adopt an outflow temperature of 40~K, while we use 80~K
for the protostellar outflow.

\subsubsection{Outflow masses}

CO is a very good outflow tracer, but in high mass star forming regions
which have massive turbulent envelopes, one can more easily see the
outflow in high density tracers like SiO or HCO$^+$. However, if the gas
density of the surrounding cloud is low, the density of the entrained
high velocity gas may not be high enough to excite HCO$^+$ in the
outflow, which is why we do not see the red-shifted outflow lobe of the
H$_2$ jet, although it is quite prominent in CO. Neither will HCO$^+$
trace high velocity gas at the very highest outflow velocities, because
the entrained gas densities become too low to excite HCO$^+$. Since most
of the mass of an outflow is at low velocities, the mass of such high
velocity gas is generally negligible compared to the total mass of the
outflow, although it can still dominate the energetics of an outflow
\citep{Mitchell91}. The uncertainty of the HCO$^+$ abundance causes much
larger uncertainties in the outflow masses derived from HCO$^+$.
However, since we know the size of the outflow from our high-spatial
resolution HCO$^+$ imaging, we can use the single dish CO \jtra21\ map
to calibrate the HCO$^+$ abundance.  Since the blue-shifted CO outflow
is anomalously weak due to self-absorption, we only use the red outflow,
and avoid near-cloud velocities, where the CO emission is optically
thick. If we integrate over the velocity range $-$46 km~s$^{-1}$ to
$-$10.5 km~s$^{-1}$, we obtain a mass of 2.3 \Msun, assuming that CO is
optically thin. Integrating HCO$^+$ over the same velocity range and
assuming that HCO$^+$ is emitted from the same gas as CO, we obtain an
HCO$^+$ abundance, [HCO$^+$/H$_2$] = 6~10$^{-8}$. This abundance is 30
times higher than the ``standard'' Orion ridge HCO$^+$ abundance,
2~10$^{-9}$ \citep{Blake87}. At extreme velocities, $-$10~km~s$^{-1}$ --
$+$20~km~s$^{-1}$, the outflow mass derived from CO is very modest, 0.3
\Msun. At near cloud velocities, $-$52~km~s$^{-1}$ -- $-$46~km~s$^{-1}$,
it is more difficult to derive a CO mass, since CO is likely to be
optically thick and the emission in this velocity range is dominated by
emission from the background cloud at $-$49~km~s$^{-1}$, see e.g.
\citep{Mitchell91}. The CO map shows that the emission from the
background cloud is approximately uniform over the map and we can
therefore correct for the emission from the cloud at  $-$49~km~s$^{-1}$,
resulting in an optically thin mass of 1.7\Msun. If we scale this mass
by the mean opacity correction factor 3.5 derived by \citet{Bontemps96}
for the sample of outflows compiled by \citet{Cabrit92}, we get an
HCO$^+$ abundance  $\sim$ 10$^{-8}$. At low velocities the HCO$^+$
abundance therefore appears to be only slightly enhanced. The
calibration of HCO$^+$ abundance relative to CO is still subject to some
uncertainty, especially since we compare high spatial resolution HCO$^+$
data to CO, which is observed with a much larger beam (22\arcsec{}). The
CO emission may therefore include emission from the H$_2$ jet and from
low density high velocity gas, which is not dense enough to be
collisionally excited in HCO$^+$. For the IRS\,9 outflow we adopt  an
[HCO$^+$/H$_2$] abundance of 10$^{-8}$ for the near-velocity gas and
6~10$^{-8}$ for high and extreme velocity gas, resulting in outflow
masses, of 8.1 \Msun\ and 7.8 \Msun\ for the red and blue outflow,
respectively (Table \ref{tbl-3}). The mass derived from HCO$^+$ is the
same in both outflow lobes. The mass discrepancy seen in earlier
observations can therefore be attributed to strong self-absorption of
the hot blue-shifted high velocity gas against colder high velocity gas,
which appears to affect CO much more strongly than HCO$^+$.

For the other two outflows we cannot use CO to calibrate the HCO$^+$
abundance, since we see none, or only small amounts of red-shifted HCO$^+$
emission, and the blue-shifted outflow lobes cannot be separated in CO.
For the mass estimates in Table \ref{tbl-3}, we have assumed that the
HCO$^+$ emission has a normal abundance ratio, [HCO$^+$/H$_2$] = 
2~10$^{-9}$. Since the blue-shifted jet overlaps with the protostellar
outflow even in our high spatial resolution HCO$^+$ observations, our
mass estimates have somewhat larger uncertainties due to the
difficulties in partitioning the high velocity emission between the
outflows. The red-shifted outflows are well separated in CO, and here we
have estimated the CO mass by spatially integrating the CO emission in
the same velocity intervals we used for estimating the mass of the
IRS\,9 outflow. We have corrected the optically thin CO mass for
near-cloud velocities with the same opacity correction (3.5) as we used
for the IRS\,9 outflow. The masses derived from CO for the red-shifted
outflows agree within a factor of  two with the blue-shifted outflow
masses derived from HCO$^+$ (Table \ref{tbl-3}), supporting our
assumption that for these outflows the HCO$^+$ abundance is not
significantly enhanced.

\section{Discussion and Conclusions}
\label{sect-discussion}

\subsection{The IRS\,9 molecular cloud is forming a cluster}

The high spatial resolution BIMA observations show that the molecular
cloud surrounding IRS\,9 breaks up into two cloud cores, one surrounding
IRS\,9, and one SE of IRS\,9. The SE cloud
core appears to be fragmenting into further sub-condensations (Section
\ref{sect-cores}). Observations of HCO$^+$ show three bipolar molecular
outflows. The dynamical time scales for each outflow is $\sim$ 10$^4$
yr, suggesting that they are all young. All three outflows have 
outflow masses in the range  20 - 60 \Msun\ and momentum fluxes $>$ 10
$^{-2}$ \Msun{} km s$^{-1}$ yr$^{-1}$, which is more than a magnitude
higher than observed for low-mass stars \citep{Cabrit92,Bontemps96}, 
but similar to values derived for outflows in other high mass star formation regions
\citep{Shepherd96,Henning00,Zhang01,Beuther02a}. Although only one of 
the stars driving the outflows have detectable radio continuum, the
outflow characteristics and high accretion rates (Section
\ref{sect-accretion}) suggest that they will all evolve into high-mass
stars.  It is very likely that there are more outflows in the IRS\,9
cloud than those  we have identified. We observe
high velocity features, which are difficult to explain with only three
outflows, and we have additionally identified an H$_2$O maser, which does
not coincide with any of the outflows. Since we have only imaged the high
velocity gas in HCO$^+$, which has a critical density of $\sim$ 10$^4$
cm$^{-3}$, we are only sensitive to very dense high velocity gas and
many outflows, especially from low mass protostars, are not always
detected in HCO$^+$. Multiple outflows near high mass protostars appear
to be common. Interferometer observations reveal multiple outflows in
several regions of massive star formation:  at least three outflows in
IRAS 05358$+$3543 \citep{Beuther02b}, as many as four outflows from the
core containing G\,35.2$-$0.7\,N \citep{Gibb03}, and seven to nine
molecular outflows in the massive star-forming region IRAS 19410$+$2336
\citep{Beuther03}. The intermediate-mass region IRAS 20293$+$3952 has
four molecular outflows, including one as highly collimated as the
jet-like outflows observed in low-mass star formation sources
\citep{Beuther04a}.

\subsection{Evidence for accretion}
\label{sect-accretion}

Radiative transfer models show that accretion onto a young stellar
object will result in self-absorbed asymmetric line profiles for
optically thick lines, with the blue-shifted side of the line being
stronger than the red-shifted one \citep{Leung77,Zhou92,Zhou93}. If the
brightness temperature of the continuum emission from the protostar
exceeds the temperature of the infalling gas, gas in front of the star
will appear in absorption, while the gas behind the star will be in
emission, resulting in an inverse P Cygni line profile, and providing strong
evidence for infall. Inverse P Cygni line profiles or line asymmetries
consistent with a centrally condensed infalling cloud have been observed
towards several massive star forming regions
\citep{Welch87,Keto88,Zhang97,Zhang98a}. In the earliest stages of the
formation of a massive protostar, the central core may be colder or
about the same temperature as the infalling envelope. The line
profiles will therefore be more symmetric or  even stronger on the
red-shifted side. However, if the massive protostar drives a hot
molecular outflow, self-absorption may be observed against the high velocity
gas.

Both IRS\,9 and the southern protostar show red-shifted self-absorption
against the hot HCO$^+$ outflows driven by these stars.  Since IRS\,9
drives an almost pole-on outflow, the outflow is expected to
largely fill in any signature of accretion, and the self-absorption in
HCO$^+$ is indeed much less pronounced than for the southern protostar
(Fig. \ref{fig-irs9-spectra}). Towards the southern protostar we observe a
deep, narrow, $\sim$ 0.8 km~s$^{-1}$ wide self-absorption feature  with
a broad, $\sim$ 4 -- 5 km~s$^{-1}$ red-shifted shoulder (Fig.
\ref{fig-irs9-spectra}). Examination of spectra and channels maps show
that the narrow absorption feature is still detectable $\sim$ 7\arcsec\
-- 10\arcsec\ from the protostar,  although it becomes even narrower in
width, 0.5 -- 0.8 km~s$^{-1}$, while the fainter broad absorption is
only detectable in the immediate vicinity of the protostar. At the position
of the protostar the broad red-shifted absorption component is
$\sim$ 8.9 K~km~s$^{-1}$~beam$^{-1}$, while the narrow component has an
integrated intensity of 1.7 --  2 K~km~s$^{-1}$~beam$^{-1}$.  Since the
FWHM of the narrow component is $\sim$ 10\arcsec, the total intensity is
7 -- 8 K~km~s$^{-1}$. The integrated intensity of the
observed self-absorption gives an estimate of the mass of  the
infalling gas. If we assume optically thin gas with a temperature of 30
K, the mass is $\sim$ 1.9  \Msun. This is a lower limit, since
the optical depth of the infalling gas must be $\ge$ 1 to be seen in
absorption. If we assume that the time scale for accretion is about the
same as for the outflow, i.e. $\sim$ 9,300 yrs, the accretion rate for
the southern protostar is  $\ge$  2 10$^{-4}$ M$_{\odot}$~yr$^{-1}$.

IRS\,9 may have an even higher accretion rate, but at near cloud
velocities the HCO$^+$ profile is completely dominated by the pole-on
outflow. At the position of IRS\,9 (Fig. \ref{fig-irs9-spectra}) there is a
narrow self-absorption feature centered on $-$53 km~s$^{-1}$, i.e.
red-shifted by $\sim$ 4 km~s$^{-1}$ from the cloud core.  Inspection of
spectra in the vicinity of IRS\,9 show that the absorption is even
broader  ($\sim$ 4 km~s$^{-1}$ wide) 5\arcsec\ to the east,
corresponding to an integrated intensity of $\sim$ 8
K~km~s$^{-1}$~beam$^{-1}$. The velocity of the self-absorption at this
position is $\sim$ $-$54 km~s$^{-1}$.  There is still  some
self-absorption 10\arcsec\ east of IRS\,9, but here the self-absorption
is narrower and less red-shifted. The self-absorption profiles suggest
higher infall velocities for IRS\,9 than for the southern protostar. The
absorption profiles also show some acceleration of the infalling gas
near IRS\,9,  but not as pronounced as towards the southern
protostar.

We can also get an estimate of the accretion rate using the observed
mass outflow rates using the arguments outlined by \citep{Beuther02a}.
\citeauthor{Beuther02a} assume that the outflows are jet-driven and that
the momentum of the outflow is equal to that of the internal jet. They
also assume that the ratio of the mass loss rate and the accretion rate
is about 0.3, and are therefore able to derive mass accretion rates from
the observed mass outflow rates. For IRS\,9 and the southern protostar
we derive mass accretion rates of 1.5 10$^{-4}$ and 7.9 10$^{-4}$
M$_{\odot}$~yr$^{-1}$. If the H$_2$ jet also originates in the IRS\,9
core, the accretion rate for IRS\,9 would be considerably higher.  Both
methods therefore suggest that the mass accretion rate is $>$ 10$^{-4}$
M$_{\odot}$~yr$^{-1}$ and is probably closer to  10$^{-3}$
M$_{\odot}$~yr$^{-1}$ for IRS\,9 and the southern protostar.

Such high accretion rates are more than sufficient to quench the
formation of an Ultra Compact \ion{H}{2} region \citep{Walmsley95}, see
also \citet{Churchwell02}, explaining why we did not detect any VLA
sources in the IRS\,9 region except IRS\,9. \citet{Anglada95} found a
good correlation between centimeter continuum emission (S$_\nu$) and
momentum flux (F$_{out}$) in low-luminosity outflow sources, with
F$_{out}$ $\propto$ S$_\nu$$^{1.1}$. Although the observed relationship
appears valid for the IRS\,9 outflow, it would predict even higher flux
densities for the H$_2$ jet and the protostellar outflow, both of which
are non-detections at 3.6 cm. Our observation suggest that this
correlation is no longer valid for young high-mass outflows.

\subsection{H$_2$O,  CH$_3$OH, and OH 1720 MHz masers}

H$_2$O masers pinpoint the location of low and intermediate luminosity
protostars, because the masers are excited in the protostellar disk or
in dense clumps in the outflow in the immediate vicinity of the star.
The IRS\,9 field contains 3 H$_2$O masers (Table \ref{tbl-2}). The first
two coincide within 1\arcsec\ with IRS\,9 and are most likely dense
clumps of shocked gas in the high velocity outflow near IRS\,9. The
third H$_2$O maser is not associated with any known source in the IRS\,9
region, nor is there any molecular outflow or shocked H$_2$ emission at
the position of the H$_2$O maser. This H$_2$O maser is probably excited
by a low-mass protostar, which is too deeply embedded to be observed in the
near-IR and too compact and faint to be detectable in dust-continuum.

The class I or class Ia methanol masers like CH$_3$OH 7$_0$ $\to$ 6$_1$
A$^+$ at 44 GHz,  are thought to be associated with outflows from
massive stars \citep{Sobolev93}. \citet{Plambeck90} carried out high
resolution studies of the CH$_3$OH 8$_0$ $\to$ 7$_1$ A$^+$ class Ia
masers in DR\,21 and DR\,21(OH) and found these masers to be located
along the interfaces between the outflows and cold dense ambient clouds.
\citet{Johnston92} mapped the CH$_3$OH 6$_2$ $\to$ 6$_1$  and 5$_2$
$\to$ 5$_1$ E transitions near 25 GHz in Orion with the VLA and found
the CH$_3$OH masers to be located near the interface of the
high-velocity outflow and the surrounding dense gas, i.e. very similar
to the results of  \citet{Plambeck90} for the DR\,21 complex.
\citeauthor{Johnston92} suggested that the methanol masers are formed
behind shock fronts, and that these shock fronts are moving at right
angles to the line of sight. The methanol masers in the IRS\,9 region
are all definitely associated with massive outflows, but the location
and radial velocities of the masers suggest that they originate in the
shearing layer between the outflow and the surrounding dense cloud,
rather than in the bow-shocks driven by the outflow. The two masers,
IRS\,9-CH$_3$OH(1) \& (2)  (Table \ref{tbl-2}) are both located in the
pole-on IRS\,9 outflow; one in the red and one in the blue outflow lobe.
The third one, IRS\,9-CH$_3$OH(3) coincides with the H$_2$ jet, also
seen as a very highly collimated outflow in HCO$^+$ (Section
\ref{sect-h2-jet}). Since their velocities are close to that of the
ambient gas, it appears that they are excited in the shearing layer
between the outflow and the surrounding molecular cloud. All class I
methanol masers in the NGC7538 molecular cloud appear to be associated
with outflows (Sandell et al, 2005, in preparation).

Most OH 1720 MHz masers are found to be coincident with \ion{H}{2}
regions, where they often coincide with OH 1665-MHz and OH 1667-MHz
masers as well as with class II methanol masers \citep{Caswell04}. OH
1720~MHz masers, however, are also known to be associated with shock
fronts from supernova remnants \citep{Wootten81,Yusef03}, and are in
rare occasions observed in shocks associated with outflows from young
stars \citep{Winnberg81,Sandell85}. We detected a new OH 1720~MHz maser
in the IRS9 field (Table ~\ref{tbl-2}). This OH 1720~MHz maser is
located in the red-shifted outflow lobe of IRS\,9 and appears to be one
of the rare OH 1720~MHz masers excited in the outflow of a young star.
Even though both class Ia CH$_3$OH masers and OH 1720~MHz masers are
found to be excited by outflows from young stars, they are not known to
be spatially coincident, suggesting that the masing conditions must be
different.

\section{Conclusion}

We have imaged the molecular cloud  surrounding IRS\,9 with
$\sim$ 5\arcsec{} angular resolution  in H$^{13}$CN \jtra10, SO 
2$_2$ $\to$ 1$_1$, HCO$^+$ \jtra10, and 3.4 mm continuum using the BIMA array at
Hat Creek, and searched for free-free emission at 8.3 GHz with 
$\sim$ 1\arcsec{} angular resolution  using the VLA. We 
supplemented these observations with VLA and JCMT archive
data. Our main results are summarized below.

1) We find that the dense gas in the IRS\,9 molecular cloud is
concentrated in two cold, massive cloud cores. The core centered on
IRS\,9 has a mass of 100 - 300 \Msun, the second core is
$\sim$20\arcsec\ SE of IRS\,9 and has a mass of $\sim$1,000 \Msun. Both
cloud cores harbor young stellar objects, which will most likely evolve
into high-mass stars.

2) We detected IRS\,9 as a weak 3.6~cm source with the VLA. 
Comparison with 6~cm VLA archive data show a flat spectrum between 6~cm
and 3.6 cm consistent with free-free emission from a collimated, ionized
jet.  We also detected IRS\,9 in dust continuum at 3.4 mm as an
extended ($\sim$15\arcsec{}) source centered on IRS\,9.

3) We resolve the previously known high velocity emission into at least
three separate bipolar outflows. IRS\,9 drives a highly collimated
extreme high-velocity outflow observed nearly pole-on. The second
outflow, the H$_2$ jet, is equally highly collimated and  jet-like and
even more massive and energetic than the IRS\,9 outflow.  This outflow
is most likely powered by an unseen object close to IRS\,9. The third
outflow is the most massive outflow in the IRS\,9 region and has an
outflow mass of $\sim$ 75 \Msun. The outflow has very high gas
densities; the outflow is observed even in H$^{13}$CN. The center of
symmetry for this outflow coincides with a strong, relatively compact
H$^{13}$CN peak in the southern core with a mass $\ge$ 250\Msun, which
appears to host a very young massive protostar.

4) Both IRS\,9 and the southern protostar show red-shifted
self-absorption profiles in HCO$^+$, consistent with high mass
accretion. The inferred mass accretion rates for both IRS\,9 and the
southern protostar are  a few times 10$^{-4}$ to $\sim$10$^{-3}$
M$_{\odot}$~yr$^{-1}$. Such high accretion rates are sufficient to
quench the formation of an \ion{H}{2} region.

5) Both IRS\,9 and the southern protostar drive hot outflows.
Examination of previously published data and single dish H$_2$CO spectra
suggest gas temperatures of 180 K for the IRS\,9 outflow and $\ge$ 80 K
for the protostellar outflow. 

6) We believe that the HCO$^+$ emission is enhanced by at least a factor
of 30 in the hot IRS\,9 outflow. The HCO$^+$ abundance may be normal in
the other two outflows.

7) The three class I methanol masers discovered in the IRS\,9 region all
coincide with massive, high velocity outflows. Since the methanol masers
have radial velocities close to the cloud velocity, it seems that they
are most likely excited in the shearing layer between the high velocity
gas and the ambient cloud. We also report the detection of an OH
1720 MHz OH maser in the IRS\,9 outflow, which is excited in conditions
similar to that of the methanol masers. The OH maser is not
spatially coincident with any of the methanol masers.

\acknowledgements 

G. S. would like to thank Ed Churchwell for useful and stimulating
discussions. The BIMA array is operated by the Universities of
California (Berkeley), Illinois, and Maryland with support from the
National Science Foundation. This work was supported in part by NSF
Grant AST 0228963 to the University of California.

{}

\clearpage

\begin{deluxetable}{lllccccr}
\tabletypesize{\scriptsize}
\tablecolumns{8}
\tablenum{1}
\tablewidth{0pt} 
\tablecaption{Continuum positions and flux densities for IRS\,9}
\label{tbl-1}
\tablehead{
\colhead{Frequency} & \colhead{$\alpha$(2000.0)} & \colhead{$\delta$(2000.0)} & \colhead{$\theta_a$ $\times$ $\theta_b$\tablenotemark{a}} & \colhead{P.A.} &\colhead{S$_{peak}$} &\colhead{S$_{int}$} & \colhead{ref}\\ 

\colhead{[GHz]} & \colhead{[h m s]} & \colhead{[$^\circ$ \arcmin\ \arcsec ]} &  \colhead{[\arcsec\ $\times$ \arcsec ]} & \colhead{[~$^\circ$~]} & \colhead{[mJy]}
& \colhead{[mJy]}  & \colhead{}}
\startdata
4.86  & 23 14 01.72 $\pm$ 0.04 &  $+$ 61 27 19.5 $\pm$ 0.3 & 2.4 $\pm$ 1.2 $\times$ 1.1 $\pm$ 1.1 & \nodata & 0.32 $\pm$ 0.11 & 1.00 $\pm$ 0.44 & 1 \\
8.46   &  23 14 01.77 $\pm$ 0.006 & $+$ 61 27 19.8 $\pm$ 0.06 & 0.7 $\pm$ 0.5 $\times$ 0.1 $\pm$ 0.1 & $+$105 $\pm$ 60 & 0.44 $\pm$ 0.06 & 0.76 $\pm$ 0.15 & 1 \\
89\tablenotemark{b} & 23 14 01.63 $\pm$ 0.05 & $+$ 61 27 20.1 $\pm$ 0.3 & 21\phantom{.} $\pm$ 3\phantom{0.} $\times$ 10\phantom{.} $\pm$ 1\phantom{0.} & \phn$+$74 $\pm$ 1\phn & \phantom{0.}18 $\pm$ 3\phantom{.00} & 85 $\pm $ 10  & 1 \\
107 & 23 14 01.78 $\pm$ 0.07 & $+$ 61 27 20.2 $\pm$ 0.3 & point-source &\nodata & \nodata & 43 & 2\\
110 & \nodata & \nodata & point-source & \nodata & \nodata & 95 & 2\\
2.2 $\mu$m\tablenotemark{c} & 23 14 01.62 $\pm$ 0.14 & $+$ 61 27 20.3 $\pm$ 1.0 & \nodata &\nodata & \nodata & 313.5 & 3\\
\enddata
\tablenotetext{a}{Deconvolved source size FWHM}
\tablenotetext{b}{The BIMA continuum image can also be fit with a two component Gaussian fit, in which case we find a unresolved point-source at the center with a flux density of 8 $\pm$ 4 mJy.}
\tablenotetext{c}{Using the 2.12 $\mu$m image of \citet{Davis98} and IRS\,10 \citep{Campbell88} as an astrometric reference gives:  23$^h$ 14$^m$ 01\psec76 $\pm$ 0\psec07, $+$ 61\degr\ 27\arcmin\ 20\ptsec5 $\pm$ 0\ptsec5.}
\tablerefs{
(1) This work; (2) van der Tak et al. (2000); (3) Tamura et al. (1991)
}

\end{deluxetable}

\clearpage

\begin{deluxetable}{llcccr}
\tablecolumns{6}
\tablenum{2}
\tablewidth{0pt} 
\tablecaption{Masers in the IRS\,9 field}
\label{tbl-2}
\tablehead{
\colhead{Name}  & \colhead{$\alpha$(2000.0)} & \colhead{$\delta$(2000.0)}  & \colhead{V$_{lsr}$} & \colhead{Flux} & \colhead{ref}\\ 
\colhead{} & \colhead{[h m s]} & \colhead{[$^\circ$ \arcmin\ \arcsec ]} & \colhead{km~s$^{-1}$} & \colhead{[Jy]}  &\colhead{}
}
\startdata
\cutinhead{H$_2$O}
IRS\,9-H$_2$O(1)             & 23 14 01.75  & $+$61 27 19.7 & $-$61.9 & 1.7 & 1\\
IRS\,9-H$_2$O(2)             & 23 14 01.72  & $+$61 27 19.9 & $-$74.3 & 3.1 & 1\\
IRS\,9-H$_2$O(3)             & 23 13  58.60 &  $+$61 27 04.5    & $-$68.8 &  16.1& 2 \\
\cutinhead{OH 1720-MHz}   
IRS\,9-OH  & 23 14 03.15  & $+$61 27 21.4 & $-$54.7      &  1.4 & 2\\
\cutinhead{CH$_3$OH 7$_0$ $\to$ 6$_1$ A$^+$} 
IRS\,9-CH$_3$OH(1)   & 23 14 01.99  &  $+$61 27 21.8    & $-$57.7  &  2.4 & 2 \\ 
IRS\,9-CH$_3$OH(2)   & 23 14 00.82  &  $+$61 27 19.4    & $-$56.3  &  1.5  & 2\\
IRS\,9-CH$_3$OH(3)   & 23 14 03.02  &  $+$61 26 46.7    & $-$57.0  &  0.70 & 2 \\
\enddata

\tablerefs{
(1) Kameya et al. (1990); (2) This work
}
\end{deluxetable}

\clearpage

\begin{deluxetable}{lrrrrrrrr}
\tablecolumns{9}
\tablenum{3}
\tablewidth{0pt} 
\tablecaption{Outflow parameters corrected for opacity and inclination. }
\label{tbl-3}
\tablehead{
\colhead{Outflow} & \colhead{t$_{dyn}$}  & \colhead{M$_{blue}$} & \colhead{M$_{red}$} & \colhead{M$_{out}$} & \colhead{P$_{out}$} &\colhead{F$_{out}$} &\colhead{E$_{kin}$} & \colhead{L$_{mec}$}\\

}
\startdata
IRS\,9\tablenotemark{a}        & 9,500   &  8 & 8  & 16 & 215 & 21 & 4 & 10\tablenotemark{b} \\
H$_2$-jet & 19,500 & 21  &10 & 31 & 620 & 30 & 14 & 29 \\
Protostar &  9,300 &  34 & 15 & 59 & 1,100 & 116 &30 &200\\
\enddata
\tablecomments{Outflow masses are computed from HCO$^+$ for blue-shifted emission and from CO for red-shifted emission (except for IRS\,9, where all parameters are derived from HCO$^+$). Masses, M$_{blue}$, M$_{red}$, and total mass, M$_{out}$, are given in [\Msun{}]. momentum P, in [\Msun{} km s$^{-1}$], momentum flux F$_{out}$, in [10$^{-3}$ \Msun{} km s$^{-1}$ yr$^{-1})$], kinetic energy E$_{kin}$, in [10$^{46}$~ergs], and mechanical luminosity, L$_{mec}$, in [\Lsun{}].}

\tablenotetext{a}{Since we used too small a bandwidth to cover the whole velocity range in HCO$^+$, we have estimated the dynamical time scale from the observed maximum CO velocity, 77 km~s$^{-1}$ and not from HCO$^+$. Outflow masses are based on HCO$^+$.}

\tablenotetext{b}{The  luminosity from extreme high velocity CO emission not seen in HCO$^+$, would add 10 \Lsun\ to the luminosity from the red-shifted outflow alone.} 
\end{deluxetable}

\clearpage

\includegraphics[ scale=1.0]{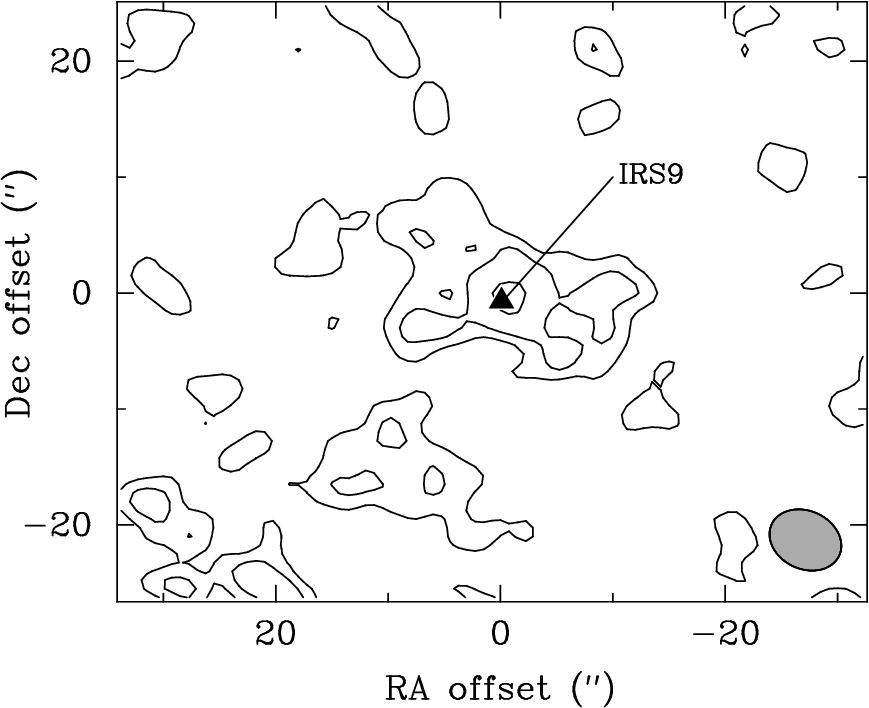}

\figcaption[]{
\label{fig-cont}
Continuum emission from IRS\,9 at 89 GHz observed with BIMA.
The lowest contour  is at 4 mJy~beam$^{-1}$ with steps of 4 mJy~beam$^{-1}$. The
position of the marginally extended (size $\le$ 1\arcsec{}) VLA source at 8.3 GHz is plotted with a filled triangle and labeled as
IRS\,9. 
The beam FWHM of BIMA is plotted in the bottom right corner of the image.}

\includegraphics[ scale=1.0]{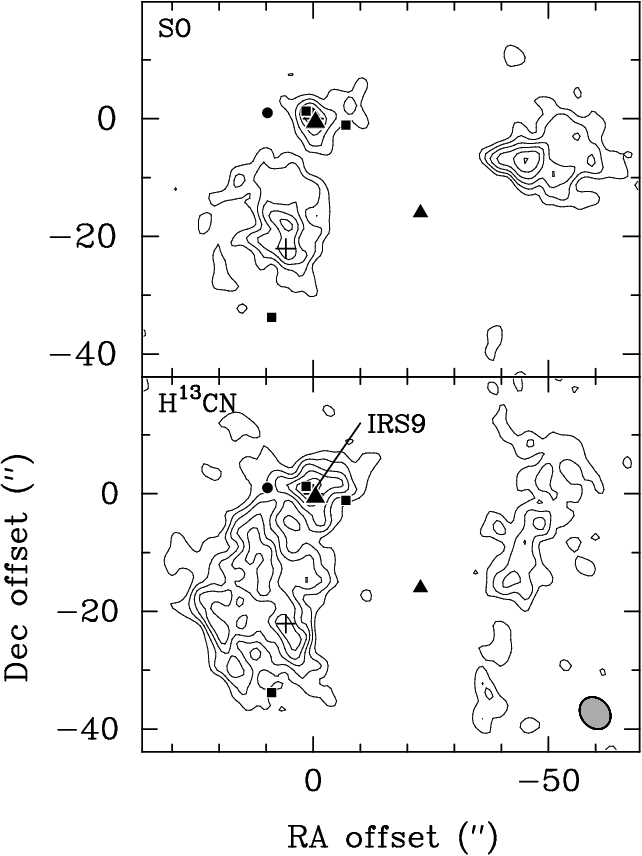}

\figcaption[]{
\label{fig-h13cn-so}
BIMA contour images of integrated H$^{13}$CN and SO emission. The offsets are
relative to the field center, \mbox{$\alpha_{2000.0}$ = 23$^h$ 14$^m$
01\psec778}, \mbox{$\delta_{2000.0}$ = $+$61\degr{} 27\arcmin{}
20\ptsec5}).  H$^{13}$CN
is integrated over the velocity interval $-$45 to $-$70 km~s$^{-1}$,
which also includes some high velocity gas (see text). SO is integrated
over a narrower velocity interval, $-$53 to $-$61 km~s$^{-1}$, to
exclude high-velocity emission. The peak flux of H$^{13}$CN and SO is
0.0854 and 0.135 Jy~km~s$^{-1}$~beam$^{-1}$, respectively.  The contour
levels start at 0.03 and 0.04 Jy~km~s$^{-1}$~beam$^{-1}$ for H$^{13}$CN
and SO, respectively.  For both molecules we plotted eight contours between
the lowest level and the peak position, The filled triangles mark known H$_2$O
masers, the filled squares are methanol class I masers, the cross marks the position of  the southern protostar, and the filled circle is the
OH 1720 MHz maser (see Table ~\ref{tbl-2}). The beam FWHM is shown in the bottom right corner of the lower panel.}

\includegraphics[ scale=0.7,angle=-90]{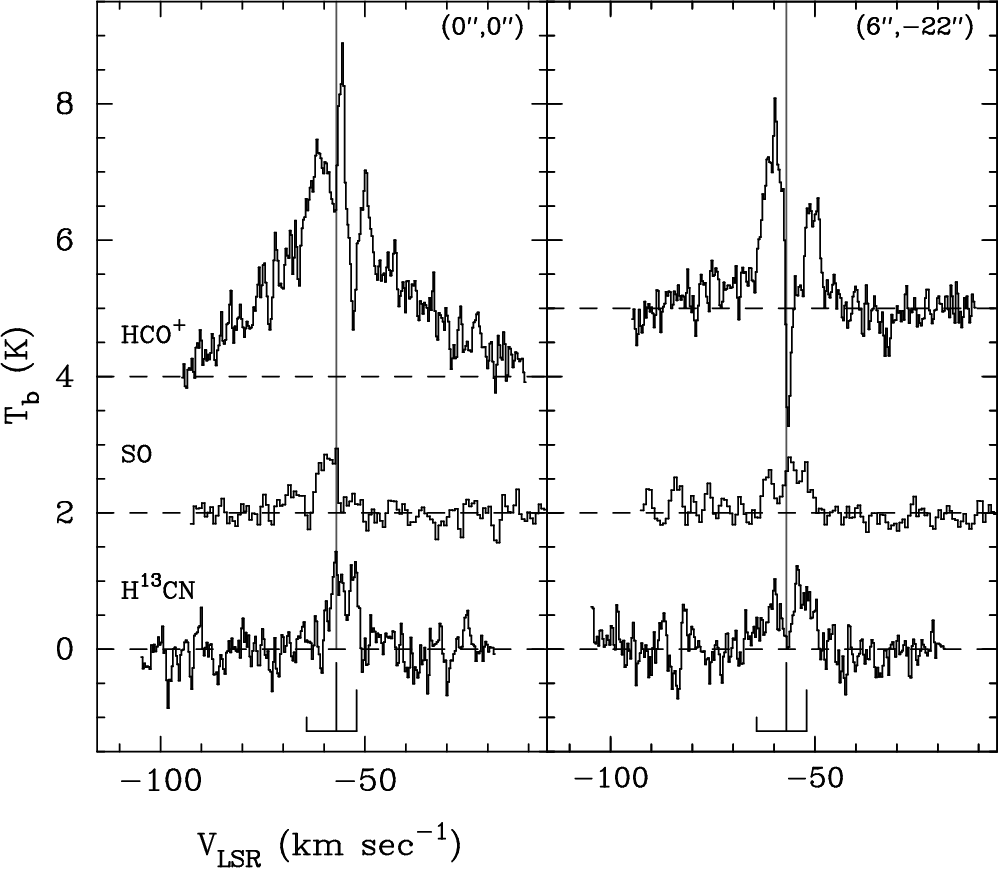}

\figcaption[]{
\label{fig-irs9-spectra} 
High resolution BIMA spectra of H$^{13}$CN, SO, and HCO$^+$ towards
IRS\,9, (0\arcsec,0\arcsec{}), and the southern protostar at
(6\arcsec\,$-$22\arcsec{}). IRS\,9 and the southern protostar show deep
red-shifted self-absorption in HCO$^+$ and evidence for high velocity
emission in H$^{13}$CN and SO. The protostar also shows deep
self-absorption in the F = $2\to1$ transition of H$^{13}$CN. The
position and relative intensity of the H$^{13}$CN hyperfine components
are indicated in the line drawings below the H$^{13}$CN spectra.}

\includegraphics[ scale=0.8,angle=-90]{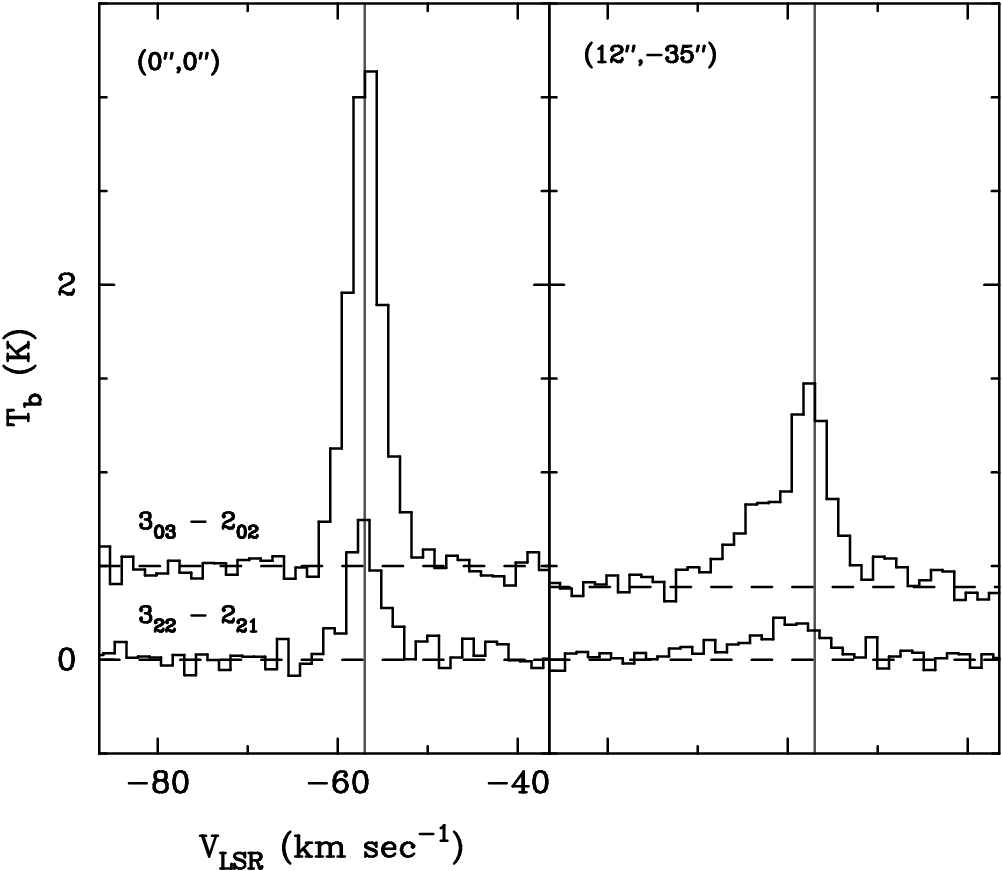}

\figcaption[]{
\label{fig-h2co} 
Single dish JCMT spectra  of  the  1.4 mm  $3_{2,2} - 2_{2,1}$ and  $3_{0,3} - 2_{0,2}$ transitions of H$_2$CO  towards IRS\,9 and a position near the southern protostar (12\arcsec,-35\arcsec{}); the latter position shows strong blue-shifted high velocity gas in both transitions. The spectra have been binned to a resolution of 1.4 km~s$^{-1}$. } 

\includegraphics[ scale=1.0]{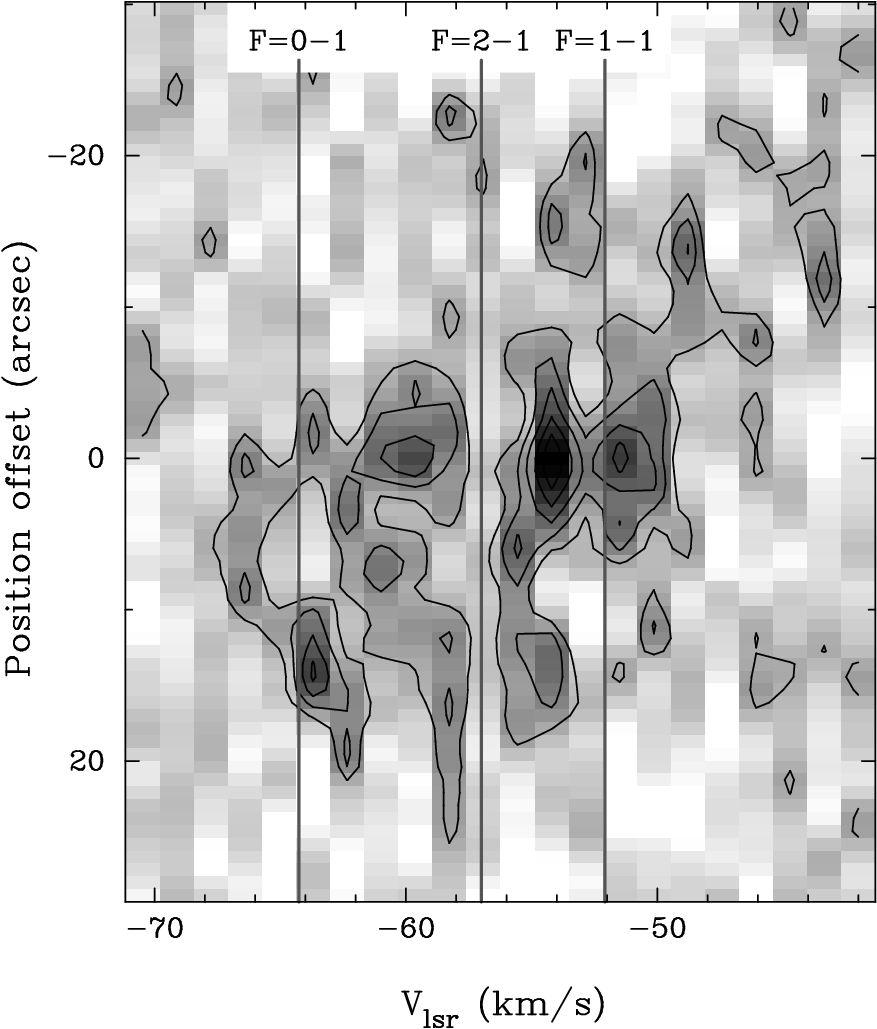}

\figcaption[]{
\label{fig-protostar}
BIMA position velocity plot of H$^{13}$CN (binned to $\Delta$V = 1.4
km~s$^{-1}$) through the protostellar core at a p.a. of 118\degr, i.e.
approximately orthogonal to the outflow powered by a protostar embedded
in this dense core. The main hyperfine component, F=$2\to1$ is
completely absorbed at the systemic velocity of the cloud core, $-57$
km~s$^{-1}$. The velocity of the three hyperfine components relative to
the systemic cloud velocity is indicated by  gray vertical  lines.
Positive offsets are to the south. }

\includegraphics[ scale=1.]{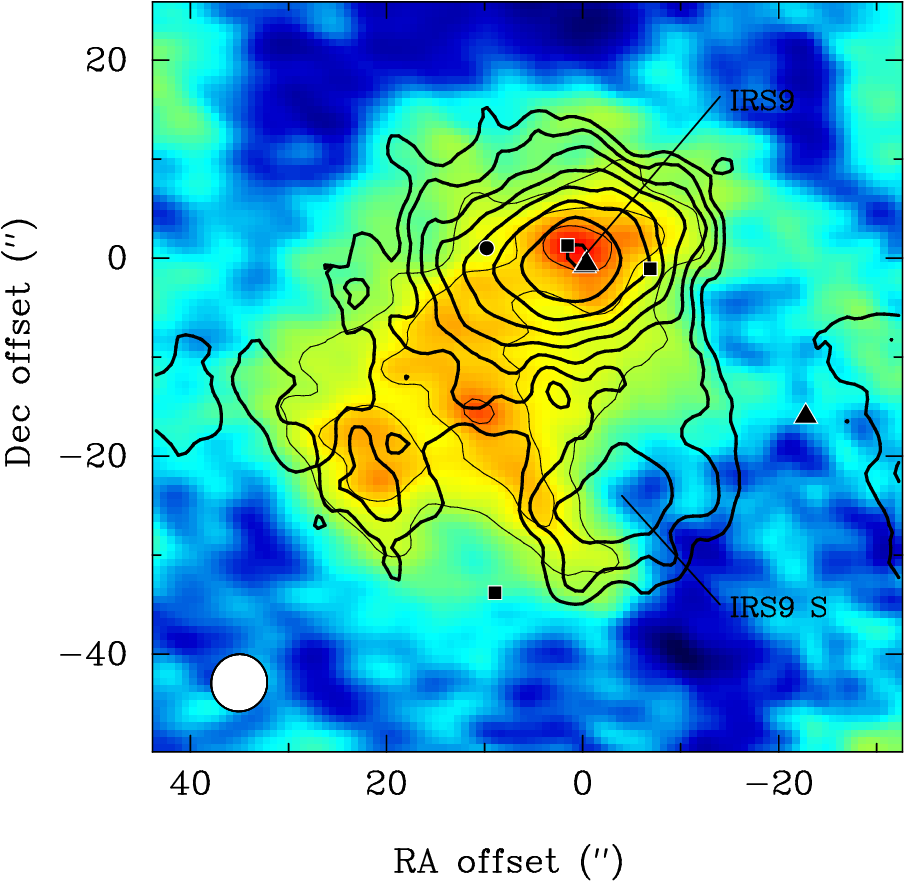}

\figcaption[]{
\label{fig-h13cn+450}
Contour plot (thick contours) of  the 450 $\mu$m SCUBA emission smoothed to
10\arcsec-resolution and overlaid on a pseudo color image of the integrated
H$^{13}$CN emission. The position of the masers are plotted as in Fig.
\ref{fig-h13cn-so}. The latter is enhanced with thin contours better to
show the extent of the H$^{13}$CN emission. The FWHM of the BIMA beam is shown in the bottom left corner of the image.}

\includegraphics[ scale=0.85]{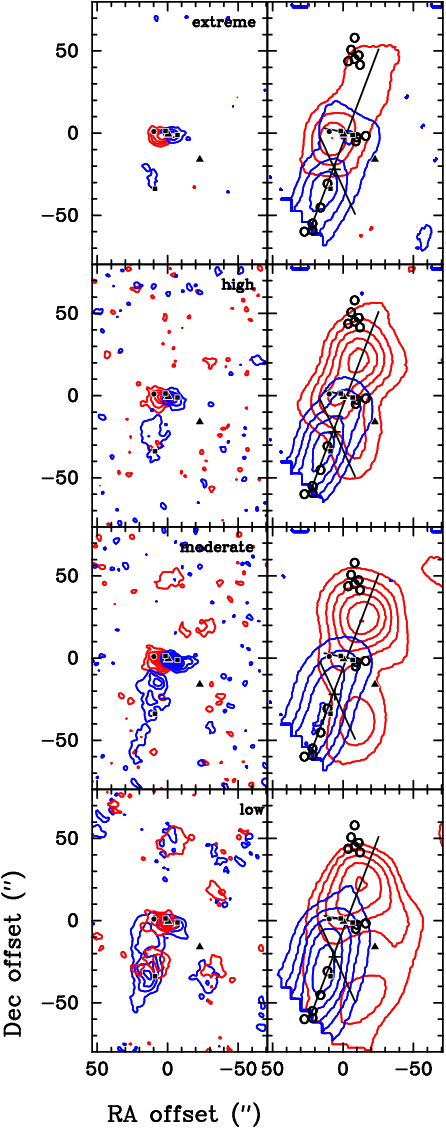}

\figcaption[]{
\label{fig-outflows}
The left panels shows contours of the high velocity emission of HCO$^+$ divided
into four velocity intervals: low, intermediate, high, and extreme, see
Section \ref{sect-flow} for details. The velocity intervals are labeled
at the top of each sub-panel. The right panels show the high velocity CO
emission over the same velocity range. The BIMA  HCO$^+$ images have an angular resolution of $\sim$ 6\arcsec, while the JCMT CO images have a resolution of $\sim$ 22\arcsec. The position of the masers are plotted as in Fig.
\ref{fig-h13cn-so}. In the right hand panel the solid lines mark the
symmetry axis of the three outflows as determined from HCO$^+$. We have
additionally marked all  shocked H$_2$ knots visible in the narrowband
H$_2$ image from \citet{Davis98}.  }

\includegraphics[ scale=1.0]{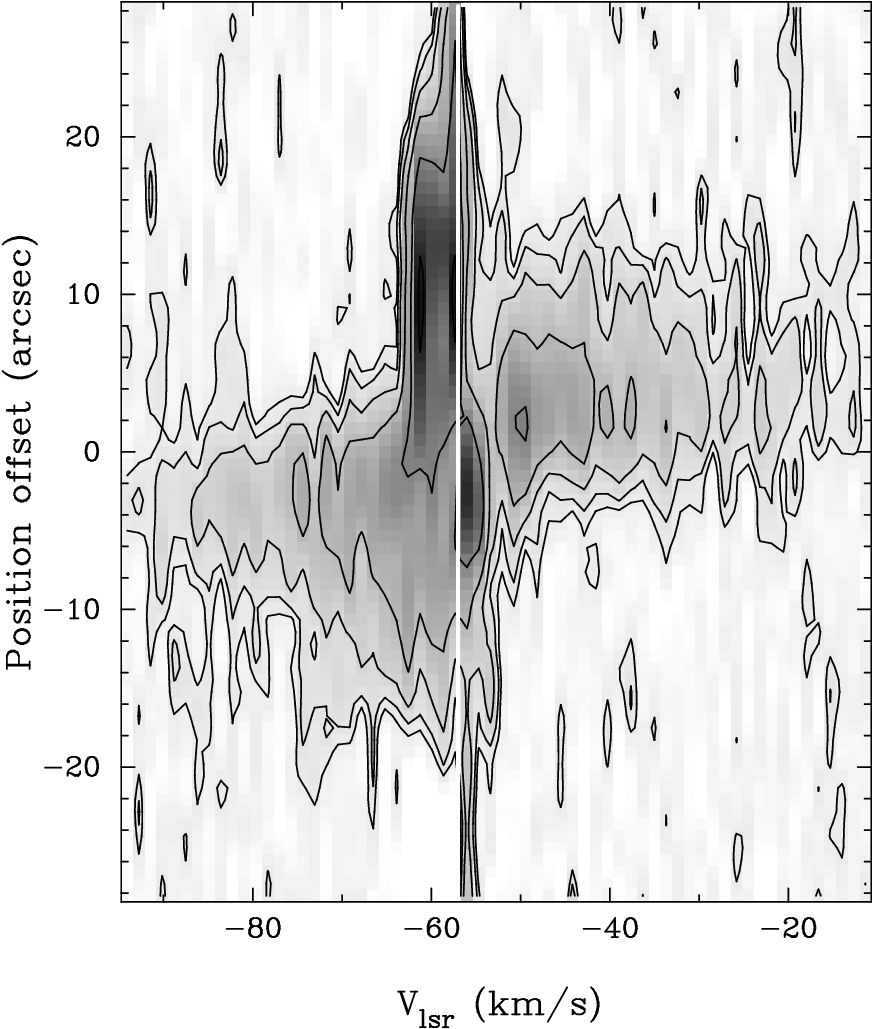}

\figcaption[]{
\label{fig-cut80}
BIMA position velocity plot of HCO$^+$ through IRS\,9 at a p.a. of 80\degr.
The systemic velocity of $-57$ km~s$^{-1}$ is indicated by the white
vertical  line.  Red-shifted self-absorption is observed over the whole outflow. To the east another cloud component at $\sim$ $-$63 km~s$^{-1}$ is detected. Positive offsets are to the east.}

\includegraphics[ scale=0.9]{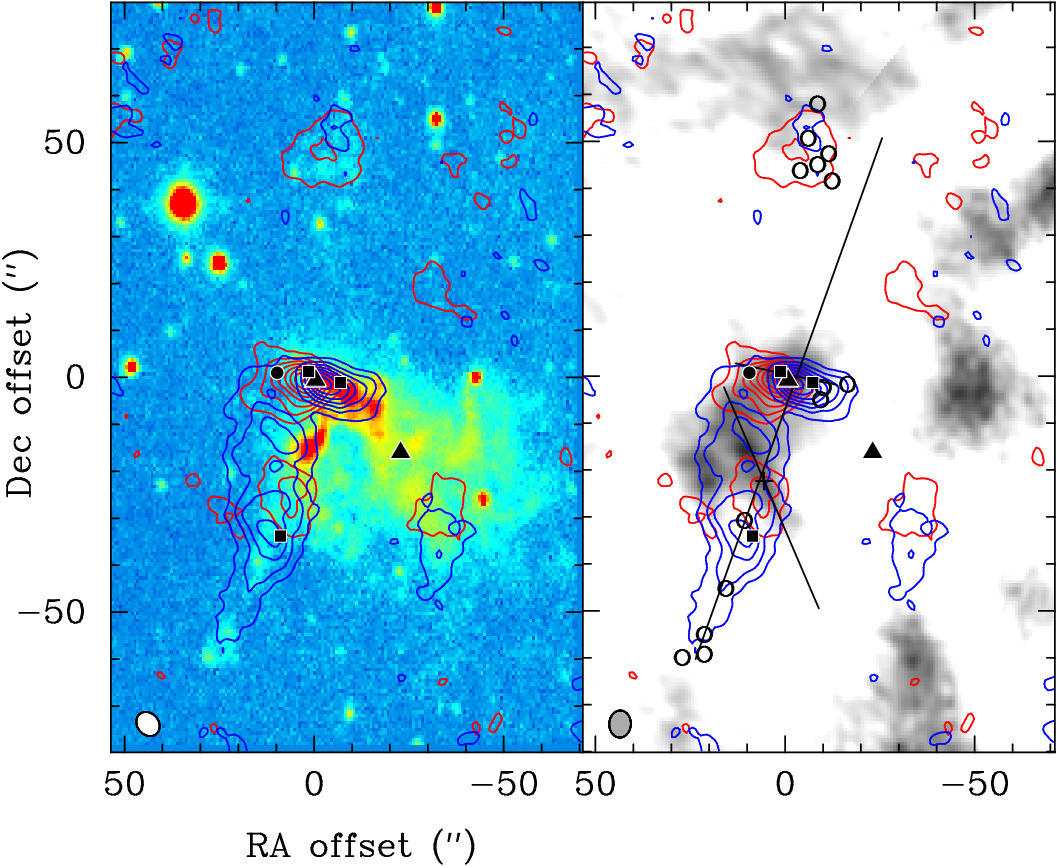}

\figcaption[]{
\label{fig-h2+hco+h13cn}
The left panel shows contour maps of HCO$^+$ integrated over low and
intermediate red and blue-shifted velocites (see text) overlaid on the
narrowband 2.12 $\mu$m H$_2$ image in color. The blue-shifted
HCO$^+$ emission follows the H$_2$ jet. About 50\arcsec\ north of IRS\,9
we observe faint red- and some blue-shifted emission coinciding with a
cluster of shocked H$_2$ knots. Another region of faint red- and
blue-shifted high velocity gas is observed $\sim$ 50\arcsec\ SW of
IRS\,9  in the southern part of the reflection nebula. The right panel
shows the same HCO$^+$ high velocity contours overlaid on the gray scale image of integrated H$^{13}$CN emission. The compact red-shifted outflow lobe
$\sim$ 25\arcsec\ south of IRS\,9 coincides with the H$^{13}$CN peak. The position of the masers are plotted as in Fig.
\ref{fig-h13cn-so} and in the rigth hand panel the open squares show the location of all shocked H$_2$ knots. The FWHM of the BIMA beam is shown in the bottom left corner, left panel: HCO$^+$, right panel: H$^{13}$CN.  }

\includegraphics[ scale=1.0]{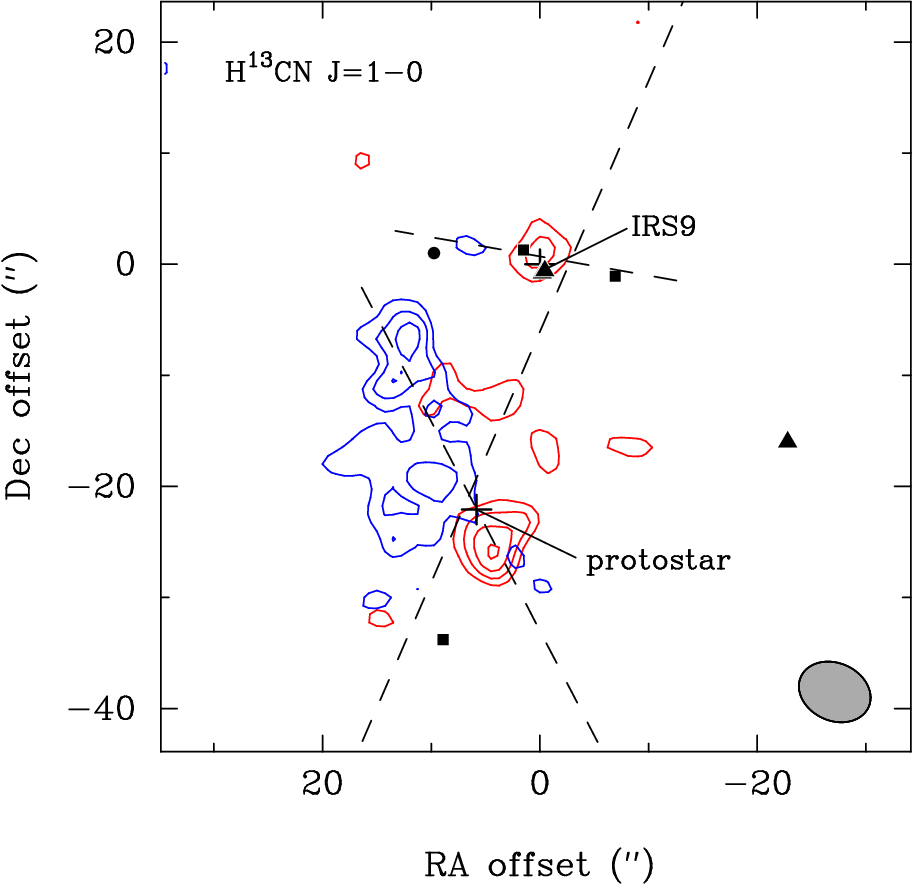}
\figcaption[]{
\label{fig-h13cn-high}
Contour image of red- and blue-shifted high velocity emission in
H$^{13}$CN \jtra10. The red-shifted emission is integrated over 10 
km~s$^{-1}$ from  $-$53.5  km~s$^{-1}$, while the blue-shifted emission
is integrated from $-$63 km~s$^{-1}$ to $-$ 69.7 km~s$^{-1}$, but
excluding a 4 km~s$^{-1}$  window around the faint F = 1 -- 1
hyperfine component. The symmetry axis and extent of the three high
velocity HCO$^+$ outflows discussed in the text are drawn by dotted
lines. The position of the masers are plotted as in Fig.
\ref{fig-h13cn-so}.  The protostellar source lies between
the red- and the blue-shifted outflow lobe of H$^{13}$CN. Faint
red-shifted emission is observed towards IRS\,9. The FWHM of the BIMA beam is shown in the bottom right corner.}


\begin{thebibliography}{}

\bibitem[Anglada(1995)]{Anglada95}
     Anglada, G.   1995,  Rev. Mex. Astron. Astrofis. Ser. de Conf., 1, 67   

 \bibitem[Beetz et al.(1976)]{Beetz76}
    Beetz, M., Els\"asser, H., Poulakos, C., \& Weinberger, R.   1976, \aap, 50, 41
    
\bibitem[Beltr\'an et al.(2004)]{Beltran04}
     Beltr\'an, M. T., Gueth, F., Guilloteau, S., \& Dutrey, A.  2004, \aap, 416,631

\bibitem[Beuther \& Schilke(2004)]{Beuther04a}
    Beuther, H., \& Schilke, P.   2004, Science, 303, 1167 
    
\bibitem[Beuther et al.(2004)]{Beuther04b}
    Beuther, H., Hunter, T. R., Zhang, Q., Sridharan, T. K., Zhao, J. -H.,
    Sollins, P., Ho, P. T. P., Ohashi, N., Su, J., Lim, J., \& Liu, S. -Y.  2004, \apj, in press      

\bibitem[Beuther, Schilke, \& Gueth(2004)]{Beuther04c}
    Beuther, H., Schilke, P., \& Gueth, F.  2004, \apj, 608, 330
              
\bibitem[Beuther, Schilke  \& Stanke(2003)]{Beuther03}
    Beuther, H., Schilke, P., \& Stanke, T.   2003, \aap, 408, 601
    
\bibitem[Beuther et al.(2002)]{Beuther02a} 
      Beuther, H., Schilke, P., Sridharan, T.~K., Menten, K.~M., Walmsley, C.~M., 
       \& Wyrowski, F.    2002a, \aap, 383, 892

\bibitem[Beuther et al.(2002)]{Beuther02b} 
      Beuther, H., Schilke, P., Gueth, F., McCaughrean, M., Andersen, M., Sridharan, T. K., 
      \& Menten, K. M.   2002b, \aap, 387, 931             

\bibitem[Blake et al.(1987)]{Blake87}
   Blake, G. A., Sutton, E. C., Masson, C. R., \& Phillips, T. G.   1987, \apj, 315, 621

\bibitem[Bohlin et al.(1987)]{Bohlin87}
     Bohlin, R. C., Savage, B. D., \& Drake, J. F.   1978, \apj, 224, 132
     
\bibitem[Bontemps et al.(1996)]{Bontemps96}
     Bontemps, S., Andr\'e, P., Terebey, S., \& Cabrit, S.  1996, \aap, 311, 858 
        
\bibitem[Cabrit \& Bertout(1992)]{Cabrit92}
    Cabrit, S., \& Bertout, C.   1992, \aap, 261, 274
    
\bibitem[Campbell \& Persson(1988)]{Campbell88}
    Campbell, B., \& Persson, S. E.    1988, \aj, 95, 1185
        
\bibitem[Caswell(2004)]{Caswell04}
    Caswell, J. L.   2004, \mnras, 349, 99
    
\bibitem[Cesaroni et al.(1997)]{Cesaroni97}
   Cesaroni, R., Felli, M., Testi, L., Walmsley, C. M., \& Olmi, L.
   1997,   \aap, 325, 725
   
\bibitem[Chernin et al.(1994)]{Chernin94}
    Chernin, L. M., Masson, C., \& Fuller, G. A.  1994, \apj, 436, 741
      
\bibitem[Chini et al.(2004)]{Chini04}  
     Chini, R., Hoffmeister, V., Kimeswenger, S., Nielbrock, M., N\"urnberger, D.
     Schmidtobreick, L., \& Sterzik, M.  2004, \nat, 429, 155

\bibitem[Churchwell(2002)]{Churchwell02}  
     Churchwell, E.  2002, \araa, 40, 27
          
\bibitem[Crampton, Georgelin \& Georgelin(1978)]{Crampton78}
     Crampton, D., Georgelin, Y. M., \& Georgelin, Y. P.  1978 \aap, 66, 1

\bibitem[Davis et al.(1998)]{Davis98}
   Davis, C.J., Moriarty-Schieven, G., Eisl\"offel, J., Hoare, M. G.,
   \& Ray, T.P.  1998, \aj, 115, 1118
   
\bibitem[Eiroa, Lenzen \& Gomez(1988)]{Eiroa88}
    Eiroa, C., Lenzen, R., \& Gomez, A. I.  1988, \aap, 190, 283
     
\bibitem[Garden \&Carlstrom(1992)]{Garden92}
    Garden, R. P., \& Carlstrom, J. E.   1992, \apj, 392, 602

\bibitem[Gibb et al.(2003)]{Gibb03}
   Gibb, A. G., Hoare, M. G., Little, L. T., \& Wright, M. C. H.  2003, \mnras, 339, 1011

\bibitem[Hasegawa \& Mitchell(1995)]{Hasegawa95}
    Hasegawa, T., \& Mitchell, G. F.   1995, \apj, 441, 665
    
\bibitem[Henning et al.(2000)]{Henning00}
     Henning, Th., Schreyer, K., Launhardt, R., \& Burkert, A.   2000, \aap, 353, 211    

\bibitem[Hildebrand(1983)]{Hildebrand83}
   Hildebrand, R. H. 1983, \qjras, 24,267
   
\bibitem[Hillenbrand(1977)]{Hillenbrand77}
    Hillenbrand, L. A.  1977, \aj, 113, 1733   
   
\bibitem[Johnston et al.(1992)]{Johnston92}
   Johnston, K. J., Gaume, R., Stolovy, S., Wilson, T. L., Walmsley, C. M., \&
    Menten, K. M.   1992, \apj, 385, 232
    
\bibitem[Kameya et al.(1989)]{Kameya89}
   Kameya, O., Hasegawa, T. I., Hirano, M., Takakubo, K. \&, Seki, M.  1989, \apj, 
   339, 222

\bibitem[Kameya et al.(1990)]{Kameya90}
   Kameya, O., Morita, K. -I., T.I., Kawabe, R., \& Ishiguro, M.  1989, \apj, 
   355, 562
  
\bibitem[Keto, Ho, \& Haschick(1988)]{Keto88} 
     Keto, E. R., Ho, P. T. P., \& Haschick, A. D.  1988, \apj, 324, 920  
   
\bibitem[Leung \& Brown(1977)]{Leung77}
     Leung, C. M., \& Brown, R. L.   1977, \apjl, 214, L73
   
\bibitem[Maeder \& Behrend(2002)]{Maeder02}
     Maeder, A., \& Behrend, R.  2002, \apss, 281, 75
   
\bibitem[Mangum \& Wootten(1993)]{Mangum93}
    Mangum, J., \& Wootten, A.   1993, \apjs, 89, 123
        
\bibitem[McKee \& Tan(2002)]{McKee02}
    McKee, C. F., \& Tan, J. C.   2002, \nat, 416, 59

\bibitem[Mitchell(1984)]{Mitchell84}
     Mitchell, G. F.   1984, \apj, 287, 665

\bibitem[Mitchell et al.(1990)]{Mitchell90}
    Mitchell, G. F., Maillard, J.-P., Allen, M., Beer, R., \&
    Belcourt, K.   1990, \apj, 363, 554

\bibitem[Mitchell \& Hasegawa(1991)]{Mitchell91}
    Mitchell, G. F.,  \& Hasegawa, T.   1991, \apjl, 371, L33
    
\bibitem[Moreno \& Chavarr\'{i}a-K(1986)]{Moreno86}
    Moreno, M. A., \& Chavarr\'ia-K, C.  1986, \aap, 161, 130
    
\bibitem[Plambeck et al.(1982)]{Plambeck82}
       Plambeck, R. L.,  Wright, M. C. H., Welch, J. H., Bieging, J. H., Baud, B.,
       Ho, P. T. P., \& Vogel, S. N.  1982, \apj, 259, 617
       
\bibitem[Plambeck \& Menten(1990)]{Plambeck90}
       Plambeck, R. L., \& Menten, K. M.  1990, \apj, 364, 555
    
\bibitem[Reynolds(1986)]{Reynolds86}
      Reynolds, S. P.   1986,  \apj, 304, 713 

\bibitem[Sandell \& Sievers(2004)]{Sandell04}
       Sandell, G., \& Sievers, A.  2004, \apj, 600, 269 

\bibitem[Sandell, Wright \& Forster(2003)]{Sandell03}
       Sandell, G., Wright, M.,  \& Forster, J. R.  2003, \apj, 590, L45   
       
\bibitem[Sandell et al.(1985)]{Sandell85}
      Sandell, G., Nyman, L. A., Haschick, A., \& Winnberg, A.   1985, in
      Lecture Notes in Physics 237, Nearby Molecular Clouds, Ed.
      G. Serra (Springer-Verlag: Berlin), p. 234
      
\bibitem[Sault, Teuben \& Wright(1995)]{Sault95}
     Sault, R. J., Teuben, P. J., \& Wright, M. C. H., 1995, in ASP Conf.
     Ser. 77: Astronomical Data Analysis Software and Systems IV, Eds.
     R.A. Shaw, H.E. Payne, and J.J.E. Hayes (Astronomical Society of the
     Pacific: San Francisco), p. 433
     
\bibitem[Schwartz \& Pratap(2002)]{Schwartz02}
      Schwartz, A., \& Pratap, P.  2002, \baas, 34, 1136
      
\bibitem[Shah \& Wootten(2001)]{Shah01}
      Shah, R. Y., \& Wootten, A.    2001, \apj, 554, 933
      
\bibitem[Shepherd \& Churchwell(1996)]{Shepherd96}
     Shepherd, D. S., \& Churchwell, E.  1996, \apj, 472, 225

     
\bibitem[Shirley et al.(2000)]{Shirley00}
     Shirley, Y. L., Evans II, N. J., Rawlings, J. M. C., \&
     Gregersen, E. M.   \apjs, 131, 249
     
\bibitem[Sobolev(1993)]{Sobolev93}
     Sobolev, A. M.   1993, Astron. Lett, 19, 293
     
\bibitem[Swade(1989)]{Swade89}
     Swade, D. A.   1989, \apj, 345, 828

\bibitem[Tamura et al.(1991)]{Tamura91}
     Tamura, M., Gatley, I., Joyce, R. R., Ueno, M., Suto, H., \& Sekiguchi, M.
     1991, \apj, 378, 611
     
\bibitem[Tielens et al.(1991)]{Tielens91}
   Tielens, A. G. G. M., Tokunaga, A. T., Geballe, T. R., \& Baas, F.  1991,
   \apj, 381, 181


\bibitem[van der Tak et al.(2000)]{Tak00}
   van der Tak, F. F. S., van Dishoeck, E. F., Evans II, N. J., \&
   Blake, G. A.  2000, \apj, 537
   
\bibitem[Walmsley(1995)]{Walmsley95}
    Walmsley, M. C.  1995,  Rev. Mex. Astron. Astrofis. Ser. de Conf., 1, 137   

\bibitem[Welch et al.(1987)]{Welch87}
    Welch, W. J., Dreher, J. W., Jackson, J. M., Terebey, S.,
     \& Vogel, S. N.  1987, Science, 238, 1550  

\bibitem[Werner et al.(1979)]{Werner79}
   Werner, M. W., Becklin, E. E., Gatley, I., Matthews, K., Neugebauer,
   G., \& Wynn-Williams, C. G. 1979, \mnras, 188, 463
   
\bibitem[Wilson \& Rood(1994)]{Wilson94}
   Wilson, T. L., \& Rood, R. T.  1994, \araa, 32, 192
   
\bibitem[Winnberg et al.(1981)]{Winnberg81}
    Winnberg, A., Graham, D., Walmsley, C. M., \& Booth, R. S.  1981, \aap, 93, 79
   
\bibitem[Wootten(1981)]{Wootten81}
    Wootten, A.    1981, \apj, 245, 105

\bibitem[Yusef-Zadeh et al.(2003)]{Yusef03}
    Yusef-Zadeh, F., Wardle, M., Rho, J., \& Sakano, M.  2003, \apj, 319
    
\bibitem[Zhang et al.(2001)]{Zhang01}
  Zhang, Q., Hunter, T. R.,  Brand, J., Sridharan, T. K., Molinari, S., Kramer, M. A.,
   \& Cesaroni, R.  2001, \apjl, 552, L167 

\bibitem[Zhang, Ho \& Ohashi(1998)]{Zhang98a}
    Zhang, Q., Ho, P. T. P., \& Ohashi, N.  1998, \apj, 494, 636 

\bibitem[Zhang, Hunter, \&Sridharan(1998)]{Zhang98}
    Zhang, Q., Hunter, T. R., \& Sridharan, T. K.  1998, \apjl, 505, L151 
   
\bibitem[Zhang \& Ho(1997)]{Zhang97}
   Zhang, Q., \& Ho, P. T. P.   1997, \apj, 488, 241
   
\bibitem[Zheng et al.(2001)]{Zheng01}
   Zheng, X. -W., Zhang, Q, Ho, P. T. P., \& Pratap, P.  2002, \apj, 550, 301 
   
\bibitem[Zhou et al.(1993)]{Zhou93} 
     Zhou, S., Evans II, N. J., K\"ompe, C., \& Walmsley, C. M.  1993, \apj, 404, 232
  
\bibitem[Zhou(1992)]{Zhou92}
     Zhou, S.  1992, \apj, 394, 204   
   
\end{thebibliography}
\end{document}